\documentclass[twocolumn,aps,pra,longbibliography,superscriptaddress,nofootinbib,10pt]{revtex4-1}
\usepackage[latin1]{inputenc}
\usepackage{graphicx,dcolumn,bm}
\usepackage{subfigure}
\usepackage{multirow}
\usepackage{array}
\usepackage{arydshln}
\usepackage{color}
\usepackage[colorlinks,linkcolor=blue,citecolor=blue,hyperindex]{hyperref}
\usepackage{amsfonts}
\usepackage{amssymb}
\usepackage{amsmath}
\usepackage{latexsym}
\usepackage{simplewick}
\usepackage{epstopdf}
\usepackage{multirow}
\usepackage{float}
\usepackage[table]{xcolor}
\usepackage{multibib}
\usepackage{mathrsfs}
\usepackage[sans]{dsfont}
\usepackage[mathscr]{eucal}

\usepackage[normalem]{ulem}
\usepackage{epsfig}
\usepackage{orcidlink}
\definecolor{Blue}{rgb}{0.0,0.0,1}
\definecolor{Red}{rgb}{1,0.0,0.0}
\definecolor{Green}{rgb}{0,0.5,0.0}
\usepackage{tikz}
\usetikzlibrary{decorations.pathmorphing}
\usetikzlibrary{arrows}
\usetikzlibrary{intersections,shapes.arrows}
\usetikzlibrary{calc}
\usetikzlibrary{quotes,angles}
\usepackage{nicefrac}
\usepackage{pgfplots}
\usepgfplotslibrary{fillbetween}
\pgfplotsset{compat=1.13,colormap={violetnew}{rgb=(0.293416, 0.0574044, 0.529412) rgb=(0.394818,0.233715,0.671945) rgb =(0.49622,0.410025,0.814477) rgb=(0.588672,0.567494,0.910066) rgb=(0.663226,0.687282,0.911765) rgb=(0.73778,0.807069,0.913465) rgb=(0.807267,0.861883,0.894034) rgb=(0.874222,0.884211,0.864039) rgb=(0.941176, 0.906538, 0.834043)}}
\usepgfplotslibrary{groupplots} 
\usepgfplotslibrary[groupplots] 
\usetikzlibrary{pgfplots.groupplots} 
\usetikzlibrary[pgfplots.groupplots] 
\usepgfplotslibrary{statistics}
\usepackage{pgfplotstable}
\tikzset{jumpdot/.style={mark=*,solid},excl/.append style={jumpdot,fill=white},incl/.append style={jumpdot,fill=black}}
\begin{document}

\title{Quantum speed limits based on Jensen-Shannon and Jeffreys divergences for general physical processes}
\author{Jucelino Ferreira de Sousa\,\orcidlink{0000-0003-0382-9781}}
\affiliation{Programa de P\'{o}s-Gradua\c{c}\~{a}o em F\'{i}sica, Universidade Federal do Maranh\~{a}o, Campus Universit\'{a}rio do Bacanga, 65080-805, S\~{a}o Lu\'{i}s, Maranh\~{a}o, Brazil}
\author{Diego Paiva Pires\,\orcidlink{0000-0003-4936-1969}}
\affiliation{Programa de P\'{o}s-Gradua\c{c}\~{a}o em F\'{i}sica, Universidade Federal do Maranh\~{a}o, Campus Universit\'{a}rio do Bacanga, 65080-805, S\~{a}o Lu\'{i}s, Maranh\~{a}o, Brazil}
\affiliation{Coordena\c{c}\~{a}o do Curso de F\'{i}sica -- Bacharelado, Universidade Federal do Maranh\~{a}o, Campus Universit\'{a}rio do Bacanga, 65080-805, S\~{a}o Lu\'{i}s, Maranh\~{a}o, Brazil}

\begin{abstract}
We discuss quantum speed limits (QSLs) for finite-dimensional quantum systems undergoing general physical processes. These QSLs were obtained using two families of entropic measures, namely the square root of the Jensen-Shannon divergence, which in turn defines a faithful distance of quantum states, and the square root of the quantum Jeffreys divergence. The results apply to both closed and open quantum systems, and are evaluated in terms of the Schatten speed of the evolved state, as well as cost functions that depend on the smallest and largest eigenvalues of both initial and instantaneous states of the quantum system. To illustrate our findings, we focus on the unitary and nonunitary dynamics of mixed single-qubit states. In the first case, we obtain speed limits \textit{\`{a} la} Mandelstam-Tamm that are inversely proportional to the variance of the Hamiltonian driving the evolution. In the second case, we set the nonunitary dynamics to be described by the noisy operations: depolarizing channel, phase damping channel, and generalized amplitude damping channel. We provide analytical results for the two entropic measures, present numerical simulations to support our results on the speed limits, comment on the tightness of the bounds, and provide a comparison with previous QSLs. Our results may find applications in the study of quantum thermodynamics, entropic uncertainty relations, and also complexity of many-body systems.
\end{abstract}

\maketitle


\section{Introduction}
\label{sec:00000000001}

Understanding entropic quantifiers has contributed to fundamental advances ranging from quantum information science to condensed-matter physics~\cite{RevModPhys.74.197}. On the one hand, entropic uncertainty relations paved the way for modern conceptual insights into Heisenberg's uncertainty principle~\cite{RevModPhys.89.015002,PhysRevLett.122.100401}, also being useful in the characterization of quantum randomness~\cite{PhysRevA.90.052327}, quantum key distribution~\cite{PhysRevLett.88.127902}, and scrambling of information~\cite{CommunPhys_2_92_2019}. On the other hand, entropic measures signal the role of entanglement in many-body quantum systems, also testifying criticality and phase transitions~\cite{RevModPhys.80.517}. In addition, entropic measures have proven useful for characterizing quantum speed limits~\cite{JMathPhys_59_012205,NewJPhys_24_065003}.

Entropic measures also have a prominent role in the subject of quantum information processing, especially with regard to the task of distinguishing quantum states. The nonuniqueness of these quantifiers manifests itself in the existence of a zoo of entropies~\cite{arXiv_1607.03104}. In this setting, quantum relative entropy (QRE) remains a paradigmatic entropic distinguishability measure of quantum states. QRE finds applications in a wide variety of physical scenarios, including the study of entanglement witnesses~\cite{PhysRevLett.103.160504}, quantum coherence measures~\cite{RevModPhys.89.041003}, asymmetry of quantum states~\cite{PhysRevA.80.012307}, and also entropy production~\cite{RevModPhys.93.035008}. Recent results address the interplay of the nonmonotonic beha\-vior of QRE and information backflow in quantum non-Markovian dynamics~\cite{PhysRevLett.127.030401,PhysRevA.106.042212}. 

Noteworthy, QRE has also been investigated within the scope of information geometry. It is a matter of fact that the Hessian matrix of QRE gives rise to the so-called Kubo-Mori inner pro\-duct~\cite{Kubo_Mori}. In this case, it is known that QRE plays a crucial role in the study of the geometry of work fluctuations in non-equilibrium quantum systems at finite temperature~\cite{PhysRevLett.123.230603,PhysRevLett.125.160602,PhysRevE.102.052117}. Furthermore, QRE provides a geometric understanding of the renormalization group for finite-dimensional quantum systems~\cite{PhysRevA.92.022330}. However, QRE is an asymmetric distinguishability measure, and is therefore not a metric in the manifold of quantum states. 

In this regard, suitable regularizations of QRE have been proposed to overcome this disadvantage, and a special role is played by the so-called quantum Jensen-Shannon divergence~\cite{PhysRevA.72.052310,PhysRevA.77.052311,PhysRevA.79.052311}, and also the quantum Jeffreys divergence~\cite{JEFF1}. These quantifiers provide u\-se\-ful distinguishability measures of two probing quantum states, although they are not true distances, and are therefore called \textit{divergences}. In fact, both measures are nonnegative, vanishing if and only if the input states are equal to each other, symmetric, but they do not sa\-tis\-fy the triangle inequality. These quantities find applications in the study of open quantum systems~\cite{PhysRevA.84.032120}, graph symmetries and quantum walks~\cite{PhysRevE.88.032806}, information processing~\cite{PhysRevLett.105.040505}, quantum coherence in multipartite systems~\cite{PhysRevLett.116.150504}, and also in the characterization of magic resource of quantum states and quantum gates for quantum computing~\cite{JPhysA_58_015303_2024}.

QRE has also been applied in the study of quantum speed limits (QSLs), i.e., the minimum time required for the evolution of a given physical process. On the one hand, Ref.~\cite{PhysRevE.103.032105} addresses QSLs for closed quantum systems by means of Petz-R\'{e}nyi and Tsallis relative entropies. On the other hand, Refs.~\cite{PhysRevA.107.052419,arXiv:2303.07415} analyzed speed limits for open quantum systems by using relative entropy of entanglement, and quantum mutual information. Recently, Ref.~\cite{ghr4-d2vb} studied a family of entropic QSLs based on $\chi$-$z$-R\'{e}nyi relative entropies that apply to ge\-ne\-ral physical processes. To the best of our knowledge, des\-pite numerous achievements regarding speed limits and relative entropies, understanding the interplay of QSLs, Jensen-Shannon divergence, and Jeffreys divergence, remains an issue to be addressed.

Here, we discuss the role of Jensen-Shannon and Jeffreys divergences in the characterization of quantum speed limits. To do so, we investigate the rate of change of these quantities and derive a class of speed limits for general physical processes. We specia\-li\-ze these results to the case of unitary dy\-na\-mics and noisy quantum channels, thus showing that the bounds depend on the smallest and largest eigenvalues of the density matrix. Importantly, our results fit into the Mandelstam-Tamm class of speed limits. In particular, we set the case of initial single-qubit states and evaluate the bounds for the depolarizing channel, phase damping channel, and generalized amplitude damping channel. We present numerical simulations to support our theoretical predictions.

We present QSLs based on entropic distinguishability measures, thereby going beyond the standard speed limits formulated in terms of quantum fidelities, such as the quantum Fisher information. Our entropic QSLs depend on the distance between two states along the quantum state space, quantified by distance measures constructed from the quantum Jeffreys divergence and the quantum Jensen-Shannon divergence. This framework ensures a faithful geometric interpretation for these entropic QSLs, particularly since the so-called quantum Jensen-Shannon distance constitutes a {\it bona fide} distance measure embedded in a metric space. This represents a significant and innovative feature compared with other entropic QSLs previously discussed in the literature. Furthermore, the proposed bounds are expected to entail low computational effort because their evaluation involves a few eigenvalues of the probe and evolved states. Consequently, our results can provide useful estimates of the speed limit time for higher-dimensional quantum systems.

Our results demonstrate a substantial advance over previous studies reported in the literature. First, our entropic QSLs encompass both unitary and nonunitary evolutions and can be applied to the study of speed limits in open quantum systems exhibiting Markovian or even non-Markovian signatures. This contrasts with Ref.~\cite{PhysRevE.103.032105}, which presented QSLs based on Petz-R\'{e}nyi and Tsallis relative entropies for closed quantum systems. Se\-cond, our QSLs explicitly account for the distinguishability between the initial and the instantaneous states of the quantum system, as quantified by both the quantum Jeffreys pseudodistance and the quantum Jensen-Shannon distance. This feature is particularly desirable for characterizing quantum speed limits for completely orthogonal states and for investigating the physical conditions under which the bound is saturated. This represents a clear advantage over Ref.~\cite{PhysRevA.106.012403}, which derived a family of QSLs based on unified quantum $(\chi,\mu)$-entropy, which depends exclusively on the sensitivity of the rate of change of the instantaneous state of the system. By contrast, our results are physically more informative. Third, our bounds require relatively low computational effort, relying on prior knowledge of the Kraus operators governing the nonunitary evolution, as well as the smallest and largest eigenvalues of the initial and instantaneous states, thereby clarifying the role of populations in the QSL. In comparison, Refs.~\cite{NewJPhys_24_065003,arXiv:2303.07415,PhysRevA.107.052419} presented QSLs that require the full spectrum of the effective Liouvillian of the open quantum system. Since our results depend on the distinguishability between the initial and final states, we expect that the entropic QSLs to be related to the amount of quantum coherences in the evolved state with respect to the eigenbasis of the initial state. This feature is not clearly evident in the results of Refs.~\cite{NewJPhys_24_065003,PhysRevA.106.012403}. Finally, we emphasize that our bounds cannot be recovered from Ref.~\cite{ghr4-d2vb} because those QSLs are formulated for $\chi$-$z$-R\'{e}nyi relative entropies with $\chi \in (0,1)$ and $1 \geq z \geq \text{max}\{\chi, 1 - \chi\}$, and $\chi \neq 1$.

The outline of the paper is as follows. In Sec.~\ref{sec:00000000002}, we review the main properties of entropic information-theoretic measures of quantum states. In Sec.~\ref{sec:00000000003}, we present upper bounds on the square roots of both Jensen-Shannon divergence and Jeffreys divergence for quantum states of a finite-dimensional quantum system undergoing a general physical process. In Sec.~\ref{sec:00000000004}, we discuss quantum speed limits based on such entropic measures. In Secs.~\ref{sec:00000000004A} and~\ref{sec:00000000004B}, we investigate these entropic QSLs for the dynamics of closed and open quantum systems, respectively. We discuss the relation between these QSLs with results previously derived in literature, also addressing the tightness of both bounds. In Sec.~\ref{sec:00000000005}, we illustrate our findings for the unitary [see Sec.~\ref{sec:00000000005A}] and nonunitary [see Sec.~\ref{sec:00000000005B}] dynamics of single-qubit states. In Sec.~\ref{sec:00000000006}, we investigate the tightness of our bounds. Finally, in Sec.~\ref{sec:00000000007} we summarize our conclusions.


\section{Entropic measures}
\label{sec:00000000002}

In this section, we briefly review the main properties of entropic information-theoretic quantifiers of quantum states. We consider a physical system described by finite-dimensional Hilbert space $\mathcal{H}$, with $\text{dim}(\mathcal{H}) = d$. The state of the system is described by a density operator $\rho \in \mathcal{S}$, where $\mathcal{S} = \{\rho \in \mathcal{H}~|~\rho^{\dagger} = \rho,~\rho \geq 0,~\text{Tr}(\rho) = 1 \}$. In this setting, given two density operators $\rho,\varrho\in\mathcal{S}$, the quantum relative entropy (QRE) is defined as~\cite{NIELSEN,Ohya_Petz}
\begin{equation}
\label{eq:0000001}
S(\rho\|\varrho) := \left\{ \begin{array}{lll}
\text{Tr}\left[\rho\left(\ln\rho - \ln\varrho\right)\right] \quad \text{if supp}(\rho) \subseteq \text{supp}(\varrho) ~, \\ 
+\infty \quad \text{otherwise} ~,
\end{array} \right.
\end{equation}
where $\text{supp}(\bullet)$ is the support of the referred density matrix. We note that Eq.~\eqref{eq:0000001} was originally introduced by Umegaki over a half century ago~\cite{RelEntr01,CommunMathPhys_33_305}, and it stands for the quantum analog of the classical Kullback-Leibler relative entropy~\cite{RelEntr02}. In particular, the former recovers the latter whenever the quantum states $\rho$ and $\varrho$ commute. We note that the QRE can be readily recast as $S(\rho\|\varrho) = - S(\rho) - \text{Tr}(\rho\ln\varrho)$, where $S(\rho) = -\text{Tr}(\rho\ln\rho)$ is the von Neumann entropy. 

Overall, QRE is (i) non-negative, i.e., $S(\rho\|\varrho) \geq 0$, with $S(\rho\|\varrho) = 0$ if and only if $\rho = \varrho$; (ii) jointly convex, i.e., ${\sum_l} \, {p_l}S({\rho_l}\|{\varrho_l}) \geq S\left({\sum_l} \, {p_l}{\rho_l}\|{\sum_l} \, {p_l}{\varrho_l}\right)$, with $0 \leq {p_l} \leq 1$ and ${\sum_l} \, {p_l} = 1$; (iii) isometric invariant, $S(V\rho{V^{\dagger}} \| V\varrho{V^{\dagger}}) = S(\rho\|\varrho)$, where ${V^{\dagger}} = {V^{-1}}$ is a unitary operator; (iv) monotonically decreasing under completely positive and trace preserving (CPTP) maps, i.e., $S\left(\mathcal{E}(\rho) \| \mathcal{E}(\varrho)\right) \leq S(\rho\|\varrho)$, with $\mathcal{E}(\bullet)$ being a given CPTP operation; (v) additive, $S({\rho_1}\otimes{\rho_2}\|{\varrho_1}\otimes{\varrho_2}) = S({\rho_1}\|{\varrho_1}) + S({\rho_2}\|{\varrho_2})$, for all ${\{{\rho_l},{\varrho_l}\}_{l = 1,2}} \in \mathcal{S}$. For more details, see Refs.~\cite{RevModPhys.74.197,arXiv:quant-ph_0004045} and references therein.

QRE satisfies several useful lower bounds and upper bounds. On the one hand, Pinsker inequality states that QRE is bounded from below by the trace distance, i.e., $S(\rho\|\varrho) \geq (1/2){\|\rho - \varrho\|_1^2}$, where ${\|A\|_1} = \text{Tr}(\sqrt{{A^{\dagger}}A})$ is the Schatten 1-norm~\cite{Ingemar_Bengtsson_Zyczkowski}. On the other hand, QRE fulfills the upper bound $S(\rho\|\varrho) \leq {\|\rho - \varrho\|_2^2}/{\kappa_{\text{min}}}(\varrho)$, with ${\|A\|_2} = \sqrt{\text{Tr}({A^{\dagger}}A)}$~\cite{JMathPhys_46_102104_2005}, while ${\kappa_{\text{min}}}(X)$ sets the smallest eigenvalue of the state $X\in\mathcal{S}$. In addition, QRE satisfies the chain of inequalities ${S_{\text{min}}}(\rho\|\varrho) \leq S(\rho\|\varrho) \leq {S_{\text{max}}}(\rho\|\varrho)$, where ${S_{\text{min}}}(\rho\|\varrho) = - \ln[\text{Tr}({\Pi_{\rho}}\varrho)]$ sets the min-relative entropy, with ${\Pi_{\rho}}$ being the projector onto $\text{supp}(\rho)$, while ${S_{\text{max}}}(\rho\|\varrho) = \ln[\text{min}\{\lambda~|~\lambda\varrho - \rho \geq 0 \}]$ defines the max-relative entropy, with $\text{supp}(\rho) \subseteq \text{supp}(\varrho)$~\cite{4957651}.

It is worth noting that quantum relative entropy is an asymmetric distinguishability measure, i.e., $S(\rho\|\varrho) \neq S(\varrho\|\rho)$, thus violating the triangular inequality. This means that QRE does not define a faithful metric over the space of quantum states. The asymmetry degree in QRE can be captured through the upper bound $|S(\rho\|\varrho) - S(\varrho\|\rho)| \leq \mathcal{G}\left(\text{min}(\kappa_{\text{min}}(\rho),\kappa_{\text{min}}(\varrho)),{\|\rho - \varrho\|_1}/2\right)$, where $\mathcal{G}(u,v) := g(u + v,u) - g(u,u + v)$ for all $-1 \leq v \leq 1$ and $\text{max}(0, -v) \leq u \leq \text{min}(1, 1 - v)$, with $g(p,q) := p\ln(p/q) + (1 - p)\ln[(1 - p)/(1 - q)]$~\cite{JMathPhys_54_073506}. To the best of our knowledge, there are two ways to overcome the lack of symmetry, thus introducing symmetric information-theoretic quantities based on QRE. 

In the following, we present two entropic measures that will be useful throughout the paper, both intrinsically related to QRE: quantum Jeffreys divergence, and quantum Jensen-Shannon divergence.


\subsection{Quantum Jeffreys divergence}
\label{sec:00000000002A}

The quantum Jeffreys divergence is defined as~\cite{JEFF1}
\begin{equation}
\label{eq:0000002}
{S_J}(\rho\|\varrho) := \frac{1}{2}[S(\rho\|\varrho) + S(\varrho\|\rho)] ~,
\end{equation}
which is based on the \textit{ad hoc} symmetrization of QRE. The subscript $J$ refers to Jeffreys. The quantum Jeffreys divergence remains finite whenever the two quantum states $\rho$ and $\varrho$ share identical supports, whereas it diverges otherwise. This follows from the fact that, to be finite, both QREs $S(\rho\|\varrho)$ and $S(\varrho\|\rho)$ in Eq.~\eqref{eq:0000002} require that $\text{supp}(\rho) \subseteq \text{supp}(\varrho)$ and $\text{supp}(\varrho) \subseteq \text{supp}(\rho)$, respectively. Importantly, Eq.~\eqref{eq:0000002} defines the quantum analog of the classical Jeffreys divergence, the latter being ori\-gi\-nally proposed about eighty years ago~\cite{doi:10.1098/rspa.1946.0056}. The factor $1/2$ is included without loss of generality. It is worth mentioning that the classical Jeffreys divergence is not a metric on the space of probabilities, nor its square root~\cite{StudSciMathHung_26_415_1991,e22020221}. For our purposes, we consider the square root of the quantum Jeffreys divergence, which reads
\begin{align}
\label{eq:0000003}
&{D_J}(\rho,\varrho) := \sqrt{{S_J}(\rho\|\varrho)} \nonumber\\
&= \sqrt{-\frac{1}{2}[S({\rho}) + S({\varrho}) + \text{Tr}({\rho}\ln{\varrho}) + \text{Tr}({\varrho}\ln{\rho})]} ~.
\end{align}
Henceforth, we refer to the distinguishability measure in Eq.~\eqref{eq:0000003} as the quantum Jeffreys pseudodistance (QJPD). Noteworthy, QJPD is known to be
\begin{enumerate}
\item[(i)] non-negative, ${D_J}(\rho,\varrho) \geq 0$, with ${D_J}(\rho,\varrho) = 0$ if and only if $\rho = \varrho$; 
\item[(ii)] symmetric, ${D_J}(\rho,\varrho) = {D_J}(\varrho,\rho)$, for all $\rho,\varrho \in \mathcal{S}$. 
\end{enumerate}
So far, there is no proof that QJPD in Eq.~\eqref{eq:0000003} defines a faithful me\-tric in the space of quantum states.


\subsection{Quantum Jensen-Shannon divergence}
\label{sec:00000000002B}
 
\begin{figure}[!t]
\begin{center}
\includegraphics[scale=0.69]{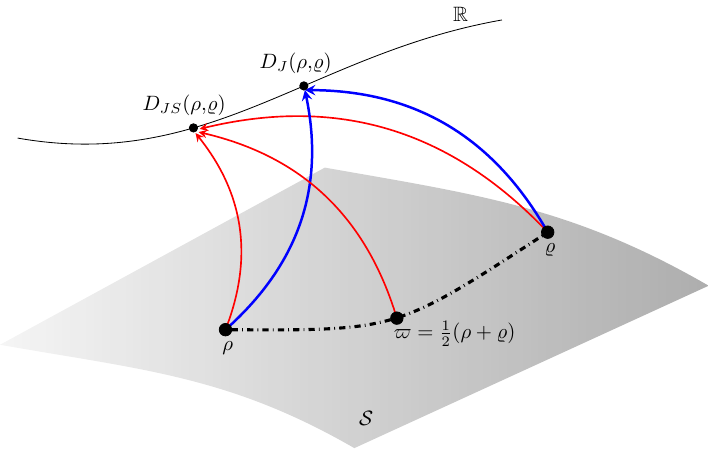}
\caption{(Color online) Overview of the role played by quantum Jeffreys pseudodistance (QJPD), ${D_J}({\rho},{\varrho})$ [see Eq.~\eqref{eq:0000003}], and quantum Jensen-Shannon distance (QJSD), ${D_{JS}}({\rho},{\varrho})$ [see Eq.~\eqref{eq:0000006}]. The QJPD (blue solid lines) and QJSD (solid red lines) map density matrices of the convex manifold of quantum states $\mathcal{S}$ (gray smooth surface) to nonnegative numbers on the real line $\mathbb{R}$ (black solid line), i.e., ${D_{J,JS}}: ({\rho},{\varrho})\in\mathcal{S} \mapsto \mathbb{R}$. Noteworty, QJSD takes into account the ``average state'' $\varpi$. QJSD is a faithful metric of quantum states, thus linking the distinguishability between $\rho$ and $\varrho$ to the length of a given path that connects these states (black dash-dotted line). QJPD, however, is not a true distance but serves as a useful entropic measure.}
\label{fig:FIG00001}
\end{center}
\end{figure}

The quantum Jensen-Shannon divergence is defined as follows~\cite{PhysRevA.72.052310,PhysRevA.77.052311,PhysRevA.79.052311}
\begin{equation}
\label{eq:0000004}
{S_{JS}}(\rho\|\varrho) := \frac{1}{2}[S(\rho\|\varpi) + S(\varrho\|\varpi)] ~,
\end{equation}
where we introduce the convex combination
\begin{equation}
\label{eq:0000005}
\varpi := \frac{1}{2}(\rho + \varrho) ~. 
\end{equation}
To clarify, the subscript $JS$ refers to Jensen-Shannon. Noteworthy, Eq.~\eqref{eq:0000004} stands as a distinguishability measure of quantum states that is based on a nontrivial symmetrization of the quantum relative entropy. To cha\-rac\-terize a well-defined information-theoretic quantity, Eq.~\eqref{eq:0000004} requires that $\text{supp}(\rho) \subseteq \text{supp}(\varpi)$ and $\text{supp}(\varrho) \subseteq \text{supp}(\varpi)$, which follows from the definition of QRE in Eq.~\eqref{eq:0000001}. We emphasize that Jensen-Shannon divergence will always remain finite since $\varpi$ is a mixed state. Indeed, it is bounded as $0 \leq {S_{JS}}(\rho\|\varrho) \leq \ln{2}$, for all $\rho,\varrho\in\mathcal{S}$, while for pure states $\rho = |{\psi}\rangle\langle{\psi}|$ and $\varrho = |{\phi}\rangle\langle{\phi}|$ it reduces to the von Neumann entropy of their convex combination, i.e., ${S_{JS}}(\rho\|\varrho) = S(\varpi)$~\cite{PhysRevA.77.052311}. Furthermore, it satisfies the property ${S_{JS}}({\rho_1}\otimes{\rho_3}\|{\rho_2}\otimes{\rho_3}) = {S_{JS}}({\rho_1}\|{\rho_2})$, for all ${\{{\rho_l}\}_{l = 1,2,3}} \in \mathcal{S}$~\cite{PhysRevA.72.052310}. Quite recently, the square root of Jensen-Shannon divergence has been proved to define a \textit{bona fide} metric over the space of quantum states~\cite{AdvMath_380_107595}. This quantity can be written as
\begin{align}
\label{eq:0000006}
{D_{JS}}(\rho,\varrho) &:= \sqrt{{S_{JS}}(\rho\|\varrho)} \nonumber\\
&= \sqrt{S\left(\varpi\right) - \frac{1}{2}\left[S\left(\rho\right)+S\left(\varrho\right)\right]} ~,
\end{align}
which will be called from now on the quantum Jensen-Shannon distance (QJSD). It has been established that the QJSD is
\begin{itemize}
\item[(i)] non-negative, ${D_{JS}}(\rho,\varrho) \geq 0$, with ${D_{JS}}(\rho,\varrho) = 0$ if and only if $\rho = \varrho$;
\item[(ii)] symmetric, ${D_{JS}}(\rho,\varrho) = {D_{JS}}(\varrho,\rho)$;
\item[(iii)] fullfils the triangular inequality, ${D_{JS}}({\rho_1},{\rho_3}) \leq {D_{JS}}({\rho_1},{\rho_2}) + {D_{JS}}({\rho_2},{\rho_3})$, for all ${\{{\rho_l}\}_{l = 1,2,3}} \in \mathcal{S}$. 
\end{itemize}
In Fig.~\ref{fig:FIG00001}, we show a schematic scenario of both the QJPD and QJSD mappings.

\section{Bounding entropic measures}
\label{sec:00000000003}

In this section, we present upper bounds on the quantum Jeffreys pseudodistance (QJPD) and the quantum Jensen-Shannon distance (QJSD) for quantum states of a physical system undergoing a general nonunitary dynamics. The system is initialized at the probe state $\rho_0 \in\mathcal{S}$, while ${\rho_t} = {\mathcal{E}_t}({\rho_0})$ describes its respective evolved state when undergoing an arbitrary, time-dependent, nonunitary dynamics generated by the completely positive and trace preserving (CPTP) map ${\mathcal{E}_t}(\bullet)$, for all $t \in [0,\tau]$. Hereafter, we set ${\hbar} = 1$. By hypothesis, the instantaneous state ${\rho_t}$ is described by a full-rank, nonsingular, and normal density matrix~\cite{RBhatia}. In this setting, it can be proved that QJPD and QJSD satisfy the upper bound
\begin{equation}
\label{eq:0000007}
{D_{J,JS}}({\rho_0},{\rho_{\tau}}) \leq \sqrt{{\int_0^{\tau}} {dt} \, {f_{J,JS}}({\rho_0},{\rho_t}) \, {\left\|\frac{d{\rho_t}}{dt}\right\|_1}} ~,
\end{equation}
where we introduce the cost function related to the QJPD, 
\begin{equation}
\label{eq:0000008}
{f_J}({\rho_0},{\rho_t}) := \frac{1}{2}\left(|\ln({\kappa_{\text{min}}}({\rho_0}) \, {\kappa_{\text{min}}}({\rho_t}))| + \frac{{\kappa_{\text{max}}}({\rho_0})}{{\kappa_{\text{min}}}({\rho_t})}\right) ~,
\end{equation}
and also the cost function related to the QJSD
\begin{equation}
\label{eq:0000009}
{f_{JS}}({\rho_0},{\rho_t}) := \frac{1}{2}\left|\ln\left({\kappa_{\text{min}}}({\rho_t})\,{\kappa_{\text{min}}}\left(\frac{{\rho_0} + {\rho_t}}{2}\right)\right)\right| ~,
\end{equation}
while ${\kappa_{\text{min/max}}}(\bullet)$ sets the smallest/largest eigenvalue of the referred density matrix, respectively. We note that Eq.~\eqref{eq:0000007} stands as the first two main results of the paper, for both the QJPD and QJSD. We find that the QJPD and QJSD are bounded above by the square root of the time-averaged of Schatten speed $\|d{\rho_t}/dt\|_1$, weighted by the auxiliary function ${f_{J,JS}}({\rho_0},{\rho_t})$. The latter plays the role of a cost function that depends on the smallest/largest eigenvalues of both initial and instantaneous states of the system. The details on the derivation of such bounds can be found in Appendix~\ref{sec:0000000000A000}. We also refer the reader to Appendixes~\ref{sec:0000000000B},~\ref{sec:0000000000C},~\ref{sec:0000000000D}, and~\ref{sec:0000000000E} for other relevant technical details.


\section{Quantum speed limits}
\label{sec:00000000004}

The quantum speed limit (QSL) establishes a lower bound on the time of evolution of quantum systems~\cite{Deffner_2017,IntJPhysB362230007}. In the last decade, QSLs have found applications in the study of optimal control~\cite{Caneva,PhysRevA.102.042606}, quantum batteries~\cite{arXiv:2006.14523,arXiv:2308.16086}, quantum correlations~\cite{EPL_62_615_2003,PhysRevA.72.032337,PhysRevA.73.049904,arXiv:2401.04599}, quantum computing~\cite{JBekenstein,Loyd}, quantum many-body systems~\cite{PhysRevLett.124.110601,PhysRevLett.126.180603,PhysRevResearch.2.032020,PhysRevA.104.052223,PhysRevX.12.011038,arXiv_2509.13519}, quantum thermodynamics~\cite{DL2010,arXiv:2204.10368,arXiv:2006.14523,arXiv:2203.12421}, non-Hermitian quantum systems~\cite{PhysRevA.104.052620,ChinPhysB_29_030304,PhysRevA.106.012403}. We note that QSLs were also addressed in the subject of quantum-to-classical transitions~\cite{PhysRevLett.120.070401,PhysRevLett.120.070402,PRXQuantum.3.020319}, and also a recent approach based on time-dependent quantum transition rates~\cite{arXiv_2506.21672}. Noteworthy, quantum speed limits have been investigated experimentally using trapped atoms~\cite{sciadvabj91192021}, nuclear spins~\cite{PhysRevA.97.052125,arXiv:2307.06558}, and also superconducting circuits~\cite{PhysRevA.110.042215}. QSLs were experimentally tested via a programmable superconducting quantum processor by emulating the dynamics of many-particle states~\cite{NatCommun_16_1255_2025}.

The works of Mandelstam and Tamm (MT)~\cite{Mandelstam1991} and Margolus and Le\-vi\-tin (ML)~\cite{1992_PhysicaD_120_188} represent a pivotal advance in the study of QSLs for closed quantum systems. These two results were unified by Levitin and Toffoli (LT)~\cite{PhysRevLett.103.160502}. On the one hand, MT showed that the unitary evolution between two quantum states $|{\psi_0}\rangle$ and $|{\psi_{\tau}}\rangle$ implies the lower bound $\tau \geq {\tau_{MT}}$, where ${\tau_{MT}} := ({\hbar}/{\Delta E})\arccos{(|\langle{\psi_0}|{\psi_{\tau}}\rangle|)}$ is the minimum time of evolution between these states~\cite{Mandelstam1991}, and $\Delta E = \sqrt{\langle{H^2}\rangle - {\langle{H}\rangle^2}}$ is the variance of a given time-independent Hamiltonian $H$ that generates the dynamics. In particular, for orthogonal states such that $\langle{\psi_0}|{\psi_{\tau}}\rangle = 0$, the QSL time becomes ${\tau_{MT}} = \pi{\hbar}/(2{\Delta E})$. On the other hand, ML proved that the minimum time of evolution for the unitary dynamics of two maximally distinguishable states satisfies the bound $\tau \geq {\tau_{ML}}$, with ${\tau_{ML}} := \hbar\pi/[2(\langle{\psi_0}|{H}|{\psi_0}\rangle - {E_0})]$~\cite{1992_PhysicaD_120_188}, where $E_0$ is the lowest energy level of the system. In turn, LT proved a tighter QSL for orthogonal states by combining MT and ML bounds as $\tau_{QSL} = \text{max}\{ {\tau_{MT}}, {\tau_{ML}}\}$~\cite{PhysRevLett.103.160502}. Recent advances involve the derivation of MT and ML bounds for pure and mixed states~\cite{PhysLettA380_689,NJPhys240550042022,arXiv:2301.10063}, correlated and separable states~\cite{PhysRevA.67.052109}, as well as the discussion on the validity of the ML bound for time-dependent Hamiltonians~\cite{JPhysA_37_L157_2004,JPhysA_51_318001_2023,PhysRevA.108.052421}. It is noteworthy the study of QSLs for unitary dynamics exploring matrix norms~\cite{PhysRevA.97.022109,arXiv:2401.01746}. We highlight Ref.~\cite{PhysRevLett.129.140403} which reported a speed limit that is dual to the ML bound, i.e., a QSL that depends on the highest energy level of the finite-dimensional quantum system.

Remarkable progress has been made in understanding QSLs in open quantum systems. Taddei {\it et al}.~\cite{Taddei2013} presented an attainable QSL bound for that depends on the quantum Fisher information. In turn, del Campo {\it et al}.~\cite{delCampoQSL} derived a quantum speed limit from the relative purity of two quantum states of a given open quantum system. Notably, these last two results provide MT-like bounds for general physical processes. In contrast, Deffner and Lutz~\cite{DLQSL} obtained an ML-type bound written in terms of the operator norm of the generator of the nonunitary dynamics. It is worth noting that speed limits strongly depend on probe states and generators of the dynamical evolution~\cite{Sun2015,Meng2015,Nicolas2016,Cianciaruso2017,Teittinen2019,Entropy_23_331_2021}. QSLs have been obtained by exploiting geometric features of the quantum state space, mainly based on contractive Riemannian metrics~\cite{DDQSL,PhysRevA.103.022210,NJPhys_24_055003_2022} and Finsler geometry~\cite{10.1142_S0129054114400073}, thus providing tighter bounds. We also mention the derivation of speed limits for open quantum systems using Schatten $p$-norms~\cite{NewJPhys_19_103018_2017,NewJPhys_21_013006_2019,PhysRevLett.126.010601,NewJPhys_24_095004_2022,PhysLettA_534_130250_2025}.

Exploring the role of entropic distinguishability measures on the derivation of QSLs may be of interest in the study of entropic uncertainty relations~\cite{RevModPhys.89.015002,PhysRevLett.122.100401,arXiv:2203.12421}. In this scenario, paradigmatic figures of merit include von Neumann entropy, quantum relative entropy, and R\'{e}nyi entropies, to name a few. On the one hand, QSLs were addressed for closed quantum systems using Petz-R\'{e}nyi and Tsallis relative entropies~\cite{PhysRevE.103.032105}. On the other hand, for open quantum systems, QSLs were investigated using negativity and quantum mutual information~\cite{PhysRevA.107.052419}, relative entropy of entanglement~\cite{arXiv:2303.07415}, unified entropies~\cite{PhysRevA.106.012403}, and most recently by means of $\chi$-$z$-R\'{e}nyi relative entropies that apply to ge\-ne\-ral physical processes~\cite{ghr4-d2vb}. Nonetheless, QSLs based on the Jensen-Shannon and Jeffreys divergences are still scarce in the literature, both for unitary or nonunitary dynamics. In the following, we derive a class of speed limits for general physical processes related to the quantum Jeffreys pseudodistance [see Eq.~\eqref{eq:0000003}], and also based on the quantum Jensen-Shannon distance [see Eq.~\eqref{eq:0000006}].

We begin with the case of the QJPD. Specifically, evaluating the time average of the right-hand side of Eq.~\eqref{eq:0000007}, the time $\tau$ required for an arbitrary nonunitary dynamics driving the state of the quantum system from $\rho_0$ to $\rho_\tau$ is lower bounded as $\tau \geq {\tau^{\text{QSL}}_J}$, with the QSL time related to the QJPD given by
\begin{equation}
\label{eq:0000010}
\mathcal{\tau}_{J}^{\text{QSL}} := \frac{{D_J^2}({\rho_0},{\rho_{\tau}})}{{\langle\langle{f_J}({\rho_0},{\rho_t}){\|d{\rho_t}/dt\|_1}\rangle\rangle_{\tau}}} ~,
\end{equation}
where $\langle\langle\bullet\rangle\rangle_{\tau} := (1/\tau){\int_0^{\tau}}\, dt \, \bullet$ stands for the time ave\-ra\-ge of a real-valued quantity. Equation~\eqref{eq:0000010} is the third result of the paper. In the following, we provide lower and upper bounds to the QSL time in Eq.~\eqref{eq:0000010}. On the one hand, Pinsker's inequality applied to the initial $\rho_0$ and final $\rho_{\tau}$ states implies that the Jeffreys divergence is lower bounded by the squared Schatten $1$-norm for these states, i.e., ${S_J}({\rho_0}\|{\rho_{\tau}}) \geq (1/2){\left\|{\rho_0} - {\rho_{\tau}}\right\|_1^2}$~\cite{Pinsker01}. In this case, one readily verifies that the QSL time in Eq.~\eqref{eq:0000010} is bounded from below as ${\mathcal{\tau}_J^{\text{QSL}}} \geq {\mathcal{\tau}_{J,\text{below}}^{\text{QSL}}}$
\begin{equation}
\label{eq:0000011}
{\mathcal{\tau}_{J,\text{below}}^{\text{QSL}}} := \frac{{\left\|{\rho_0} - {\rho_{\tau}}\right\|_1^2}}{2{\langle\langle{f_J}({\rho_0},{\rho_t}){\|d{\rho_t}/dt\|_1}\rangle\rangle_{\tau}}} ~,
\end{equation}
where the trace distance signals how far apart the two states $\rho_0$ and $\rho_{\tau}$ are in the space of quantum states. On the other hand, since $S({\rho_{0,\tau}}\|{\rho_{\tau,0}}) \leq {\|{\rho_0} - {\rho_{\tau}}\|_2^2}/{\kappa_{\text{min}}}({\rho_{\tau,0}})$, we conclude that the QSL time in Eq.~\eqref{eq:0000010} satisfies the upper bound ${\mathcal{\tau}_J^{\text{QSL}}} \leq {\mathcal{\tau}_{J,\text{above}}^{\text{QSL}}}$
\begin{equation}
\label{eq:0000012}
{\mathcal{\tau}_{J,\text{above}}^{\text{QSL}}} := \frac{({\kappa_{\text{min}}}({\rho_0}) + {\kappa_{\text{min}}}({\rho_{\tau}})) \, {\left\|{\rho_0} - {\rho_{\tau}}\right\|_2^2}}{2{\langle\langle{f_J}({\rho_0},{\rho_t}){\|d{\rho_t}/dt\|_1}\rangle\rangle_{\tau}} \,  {\kappa_{\text{min}}}({\rho_0}) \, {\kappa_{\text{min}}}({\rho_{\tau}})} ~,
\end{equation}
which in turn depends on the Schatten $2$-norm. It turns out that by combining Eqs.~\eqref{eq:0000011} and~\eqref{eq:0000012}, the hierarchical bound ${\mathcal{\tau}_{J,\text{below}}^{\text{QSL}}} \leq {\mathcal{\tau}_J^{\text{QSL}}} \leq {\mathcal{\tau}_{J,\text{above}}^{\text{QSL}}}$ is found, which restricts the possible values of the QSL time based on QJPD.

Next, we address entropic speed limits based on QJSD, thus taking advantage from the bound in Eq.~\eqref{eq:0000007}. The time required for an arbitrary nonunitary dynamics driving the state of the quantum system from ${\rho_0}$ to ${\rho_{\tau}}$ is lower bounded as $\tau\geq\tau_{JS}^{\text{QSL}}$, where we define the entropic QSL time related to QJSD as follows
\begin{equation}
\label{eq:0000013}
\mathcal{\tau}_{JS}^{\text{QSL}} := \frac{{D_{JS}^2}({\rho_0},{\rho_{\tau}})}{{\langle\langle{f_{JS}}({\rho_0},{\rho_t}){\|d{\rho_t}/dt\|_1}\rangle\rangle_{\tau}}} ~.
\end{equation}
Equation~\eqref{eq:0000013} is the fourth result of the paper. The QSL time in Eq.~\eqref{eq:0000013} scales with the inverse of the time average of the Schatten speed, weighted by the cost function. It also depends on the quantum Jensen-Shannon distance that characterizes the distinguishability between the initial $\rho_0$ and final $\rho_{\tau}$ states. The bound requires a few quantities of the system, namely, the smallest eigenvalues of the initial and instantaneous states.


\subsection{QSL\lowercase{s} and entropic measures in closed quantum systems}
\label{sec:00000000004A}

Here, we specialize the QSL bounds related to the QJPD [see Eq.~\eqref{eq:0000010}], and also the QJSD [see Eq.~\eqref{eq:0000013}] for the unitary dynamics of finite-dimensional closed quantum systems. We consider the dynamics of the closed system is governed by a time-independent Hamiltonian $H$, and the evolved state of the system is given by ${\rho_t} = {U_t}{\rho_0}{U_t^{\dagger}}$, with ${U_t} = {e^{-i t H}}$ being the evolution operator. In this case, the Schatten speed reads ${\|d{\rho_t}/dt\|_1} = {\|(-i)[H,{\rho_0}]\|_1}$.

For the QJPD, the speed limit time in Eq.~\eqref{eq:0000010} becomes
\begin{equation}
\label{eq:0000014}
{\mathcal{\tau}_J^{\text{QSL}}} = \frac{{D_J^2}({\rho_0},{\rho_{\tau}})}{(|\ln(\kappa_{\text{min}}({\rho_0}))| + \eta({\rho_0})/2) \, {\|(-i)[H,{\rho_0}]\|_1}} ~,
\end{equation}
where $\eta({\rho_0}) := {\kappa_{\text{max}}}({\rho_0})/{\kappa_{\text{min}}}(\rho_{0})$ stands for the condition number of the density matrix~\cite{Bathia}. It is worth noting that the eigenvalues of the density matrix remain unchanged under unitary evolutions, i.e., ${\kappa_{\text{min}/\text{max}}}({\rho_t}) = {\kappa_{\text{min}/\text{max}}}({\rho_0})$, for all $0 \leq t \leq \tau$. For the QJSD, the QSL bound in Eq.~\eqref{eq:0000013} yields
\begin{equation}
\label{eq:0000015}
{\mathcal{\tau}_{JS}^{\text{QSL}}} := \frac{2{D_{JS}^2}({\rho_0},{\rho_{\tau}})}{{\langle\langle \left|\ln\left(\kappa_{\text{min}}({\rho_0})\,{\kappa_{\text{min}}}\left(\frac{{\rho_0} + {\rho_t}}{2}\right)\right)\right|\rangle\rangle_{\tau}} {\|(-i)[H,{\rho_0}]\|_1} } ~,
\end{equation}
where we have used the fact that the convex sum ${\varpi_t} = (1/2)({\rho_0} + {\rho_t})$ exhibits time-dependent eigenvalues, since $\rho_0$ and $\rho_t$ are noncommuting density matrices.

Equations~\eqref{eq:0000014} and~\eqref{eq:0000015} represent the fifth result of the paper. They fit into the class of Mandelstam-Tamm speed limits. To see this point, we note that $ {\|(-i)[H,{\rho_0}]\|_1} \leq 2 {\Delta{H}}$, where ${(\Delta{H})^2} = \text{Tr}({\rho_0}{H^2}) - {[\text{Tr}({\rho_0}H)]^2}$ is the variance of the Hamiltonian $H$~\cite{boundFisher,SYu_1302.5311,PhysRevA.87.032324,GToth_1701.07461}. Hence, we have that the entropic QSLs are bounded from below as ${\mathcal{\tau}_J^{\text{QSL}}} \geq {D_J^2}({\rho_0},{\rho_{\tau}})/[(2|\ln(\kappa_{\text{min}}({\rho_0}))| + \eta({\rho_0})) \Delta{H}]$, and ${\mathcal{\tau}_{JS}^{\text{QSL}}} \geq {D_{JS}^2}({\rho_0},{\rho_{\tau}})/[{\langle\langle \left|\ln\left(\kappa_{\text{min}}({\rho_0})\,{\kappa_{\text{min}}}\left(\frac{{\rho_0} + {\rho_t}}{2}\right)\right)\right|\rangle\rangle_{\tau}} \Delta{H}]$, both of them related to the fluctuations of the generator of the dynamics.

It is worth noting that Ref.~\cite{PhysRevE.103.032105} reported a speed limit for closed quantum systems using symmetrized relative entropy. In detail, such a bound was obtained via Cauchy-Schwarz inequality, and depends on the Schatten $2$-norm of the commutator of the Hamiltonian and the probe/evolved state of the system, and also the Schatten $2$-norm of the logarithm of the initial state. However, Eq.~\eqref{eq:0000014} is expected to be tighter since it was derived via H\"{o}lder's inequality related to Schatten $1$-norm and operator norm. In addition to the squared QJPD, i.e., the symmetrized relative entropy, Eq.~\eqref{eq:0000014} depends on the smallest and largest eigenvalues of the probe state, and also on the Schatten $1$-norm of the commutator of the Hamiltonian and the initial state of the system.


\subsection{QSL\lowercase{s} and entropic measures in open quantum systems}
\label{sec:00000000004B}

Here, we illustrate the results in Eqs.~\eqref{eq:0000010} and \eqref{eq:0000013} for the nonunitary dynamics governed by the completely positive and trace-preserving (CPTP) map, ${\mathcal{E}_t}(\bullet) = {\sum_j}{K_j} \bullet {K_j^{\dagger}}$, with $\{{K_j}\}_{j = 1,...,s}$ being the set of time-dependent Kraus operators, with ${\sum_j}{K_j^{\dagger}}{K_j} = \mathbb{I}$~\cite{NIELSEN}. In this framework, by applying the triangular ine\-qua\-li\-ty ${\|{\sum_j}{\mathcal{O}_j}\|_1} \leq {\sum_j}{\|{\mathcal{O}_j}\|_1}$, and also using the invariance of the Schatten norm ${\|{\mathcal{O}^{\dagger}}\|_1} = {\|\mathcal{O}\|_1}$, one obtains the upper bound on the Schatten speed as $\left\|{d{\rho_t}}/{dt}\right\|_1 \leq 2{\sum_j}{\|{K_j}{\rho_0}{(d{K}_j^{\dagger}/{dt})}\|_1}$. Therefore, by substituting this result in Eqs.~\eqref{eq:0000010} and \eqref{eq:0000013}, one gets the QSL time based on the QJPD as
\begin{equation}
\label{eq:0000016}
\mathcal{\tau}_J^{\text{QSL}} = \frac{{D_J^2}({\rho_0},{\rho_{\tau}})}{2 \, {\langle\langle{f_J}({\rho_0},{\rho_t}) \, {\sum_j}{\|{K_j}{\rho_0}{(d{K}_j^{\dagger}/{dt})}\|_1} \rangle\rangle_{\tau}}} ~,
\end{equation}
and also the QSL time based on the QJSD as
\begin{equation}
\label{eq:0000017}
{\mathcal{\tau}_{JS}^{\text{QSL}}} = \frac{{D_{JS}^2}({\rho_0},{\rho_{\tau}})}{2\, {\langle\langle{f_{JS}}({\rho_0},{\rho_t}) \, {\sum_j}{\|{K_j}{\rho_0}{(d{K}_j^{\dagger}/{dt})}\|_1}\rangle\rangle_{\tau}}} ~,
\end{equation}
with the cost functions ${f_{J,JS}}({\rho_0},{\rho_t})$ given in Eqs.~\eqref{eq:0000008} and~\eqref{eq:0000009}, respectively.

The entropic QSL bounds in Eqs.~\eqref{eq:0000016} and~\eqref{eq:0000017} describe the sixth result of the paper. They are characterized by the smallest/largest eigenvalues of probe and evolved states that are encoded in each of the cost functions, and also by the set of Kraus operators that describes the effective nonunitary dynamics of the finite-dimensional quantum system. Furthermore, it turns out that ${\mathcal{\tau}_{J}^{\text{QSL}}}$ and ${\mathcal{\tau}_{JS}^{\text{QSL}}}$ depend on the QJPD and QJSD, respectively, both entropic quantities capturing the distinguishability between initial ${\rho_0}$ and final ${\rho_{\tau}}$ states of the dynamics.

In the following, we compare Eqs.~\eqref{eq:0000016} and~\eqref{eq:0000017} with previous QSLs addressed in the literature. On the one hand, Refs.~\cite{NewJPhys_24_065003,arXiv:2303.07415,PhysRevA.107.052419} discussed speed limits based on relative entropies of entanglement that depend on prior knowledge of the full spectrum of the Liouvillians that governs the nonunitary dynamics. On the other hand, Ref.~\cite{PhysRevA.106.012403} presents QSLs based on unified entropies for general physical processes. This family of entropies does not encompass QJPD nor QJSD as particular cases. Next, Ref.~\cite{ghr4-d2vb} reported entropic QSLs based on $\chi$-$z$-R\'{e}nyi relative entropies, the latter recovering Umegaki's relative entropy by fixing $z = 1$ and taking $\chi \rightarrow 1$. However, we note that such speed limits are valid only for $\chi \in (0,1)$ and $\text{max}\{\chi,1 - \chi\} \leq z \leq 1$, which implies that they can not provide QSLs related to standard quantum relative entropy. Last but not least, since a different combination of inequalities was used to derive Eqs.~\eqref{eq:0000016} and~\eqref{eq:0000017}, one expects that another type of speed limit would be obtained when compared with the results in Ref.~\cite{ghr4-d2vb}.


\section{Examples}
\label{sec:00000000005}

To illustrate our findings, we consider a two-level system initialized at the mixed single-qubit state ${\rho_0} = (1/2)\left(\mathbb{I} + \vec{r}\cdot\vec{\sigma}\right)$, where $\vec{r} = r \hat{r}$ is the Bloch vector, with $\hat{r} = \left(\sin\theta\cos\phi,\sin\theta\sin\phi,\cos\theta\right)$, where $0 \leq r < 1$, $0 \leq \theta \leq \pi$, and $0 \leq \phi < 2\pi$, while $\vec{\sigma} = \left({\sigma_x},{\sigma_y},{\sigma_z}\right)$ is the vector of Pauli matrices, and $\mathbb{I}$ is the $2\times2$ identity matrix. The initial state exhibits the eigenvalues $\kappa_{\text{max}/\text{min}}({\rho_0}) = (1/2)(1 \pm r)$. On the one hand, we investigate the entropic QSLs for unitary evolutions generated by time-independent Hamiltonians [see Sec.~\ref{sec:00000000005A}]. On the other hand, we address the QSLs related to nonunitary evolutions of single-qubit states in terms of some prototypical quantum channels [see Sec.~\ref{sec:00000000005B}].


\subsection{Unitary evolution}
\label{sec:00000000005A}

Let $H = \vec{n}\cdot\vec{\sigma}$ be a time-independent Hamiltonian dri\-ving the unitary evolution of the prototype two-level system, where $\vec{n} = \left({n_x},{n_y},{n_z}\right)$ is a three-dimensional vector. We note that ${\|(-i) [H,\rho_{0}]\|_1} = 2 r \|\vec{n}\| \|\hat{n}\times\hat{r}\,\|$, where $\hat{n} = \vec{n}/\|\vec{n}\|$ is a unit vector, with ${\|\vec{n}\|^2} = {n_x^2} + {n_y^2} + {n_z^2}$ being the squared Euclidean norm. The ins\-tan\-ta\-neous state of the system is given by ${\rho^C_t} = {U_t}{\rho_0}{U_t^{\dagger}} = (1/2)(\mathbb{I} + {\vec{r}^{\,C}_t}\cdot\vec{\sigma})$, where ${U_t} = {e^{- i t H}} = \cos(\|\vec{n}\|{t})\mathbb{I} - i\sin(\|\vec{n}\|{t})(\hat{n}\cdot\vec{\sigma})$ is the unitary evolution o\-pe\-ra\-tor, with ${U_t}{U_t^{\dagger}} = {U_t^{\dagger}}{U_t} =\mathbb{I}$. In turn, the convex combination ${\varpi^C_t} = (1/2)({\rho_0} + {\rho^C_t}) = (1/2)(\mathbb{I} + (1/2)(\vec{r} + {\vec{r}^{\,C}_t})\cdot\vec{\sigma})$ displays the eigenvalues $\kappa_{\text{max}/\text{min}}({\varpi^C_t}) = (1/2)(1 \pm {\nu^C_t})$, where we define
\begin{equation}
\label{eq:0000018}
{\nu^C_t} := r\sqrt{1 - {\|\hat{n}\times\hat{r}\|^2} \, {\sin^2}(\|\vec{n}\|t)} ~.
\end{equation}

Next, we address the entropic QSL time based on the QJPD [see Eq.~\eqref{eq:0000014}]. To do so, we first evaluate the QJPD, thus finding that ${D_J}({\rho_0},{\rho^C_{\tau}}) = \sqrt{S({\rho_0}\|{\rho^C_\tau})}$, where the Umegaki relative entropy for single-qubit states is given by [see Appendix~\ref{sec:0000000000E}]
\begin{equation}
\label{eq:0000019}
S({\rho_0}\|{\rho^C_{\tau}}) = r\ln\left(\frac{1 + r}{1 - r}\right){\|\hat{n}\times\hat{r}\|^2}\, {\sin^2}(\|\vec{n}\|\tau) ~.
\end{equation}
Hence, the time required for a unitary process driving the state of the quantum system from ${\rho_0}$ to ${\rho_{\tau}}$ is lower bounded as $\tau\geq\tau_{{J}}^{\text{QSL}}$, where
\begin{equation}
\label{eq:0000020}
{\tau_J^{\text{QSL}}} = \frac{\|\hat{n}\times\hat{r}\| \, \ln\left(\frac{1 + r}{1 - r}\right){\sin^2}(\|\vec{n}\|\tau)} {\|\vec{n}\|\left(2\left|\ln\left(\frac{1 - r}{2}\right)\right| + \frac{1 + r}{1 - r}\right)} ~.
\end{equation}
It is worth noting that ${\tau_J^{\text{QSL}}}$ in Eq.~\eqref{eq:0000020} is a harmonic function of the dimensionless parameter $\|\vec{n}\|\tau$, thus going to zero for all $\|\vec{n}\|\tau = k\pi$, with $k \in \mathbb{N}$. In particular, for $\|\vec{n}\|\tau \ll 1$, it approaches the value 
\begin{equation}
\label{eq:0000021}
{\tau_J^{\text{QSL}}} \approx \frac{\|\hat{n}\times\hat{r}\| \, \|\vec{n}\|{\tau^2} \, \ln\left(\frac{1 + r}{1 - r}\right)}{2\left|\ln\left(\frac{1 - r}{2}\right)\right| + \frac{1 + r}{1 - r} } ~, 
\end{equation}
and grows quadratically with time $\tau$. The entropic QSL time in Eq.~\eqref{eq:0000020} is sensitive to the role played by the coherences of the probe state with respect to the energy eigenstates of the Hamiltonian. To see this point, one verifies that $\|\hat{n}\times\hat{r}\| = \sqrt{1 - {(\hat{n}\cdot\hat{r})^2}} = \sqrt{2\,{\mathcal{I}_H}({\rho_0})}/({\|\vec{n}\|}r)$, where ${\mathcal{I}_H}({\rho_0}) = -(1/4)\text{Tr}({[{\rho_0},H]^2})$ is a coherence measure~\cite{PhysRevLett.113.170401}. This means that the minimum time it takes to change the state of the system will be influenced by the amount of coherences stored in its initial state. We find that ${\tau_J^{\text{QSL}}}$ in Eq.~\eqref{eq:0000020} vanishes for probe states that are incoherent with respect to the eigenbasis of $H$, i.e., one gets ${\mathcal{\tau}_J^{\text{QSL}}} = 0$ for parallel unit vectors $\hat{r}$ and $\hat{n}$.

\begin{table*}[!t]
\caption{Information-theoretic quantities related to the entropic QSLs for single-qubit states evolving under the quantum channels depolarizing, phase damping, and generalized amplitude damping channel. Here, $\lambda_t = 1 - {e^{-\Gamma t}}$, with $\Gamma$ being the decoherence rate. We also define ${c_{s\alpha}} = (1/2)(1 + {(-1)^s}) + {(-1)^{s - 1}}\alpha$, with $s = \{1,2\}$, where $\alpha \in [0,1]$ is a dimensionless parameter that encodes the temperature effects in the noisy channel.}
\begin{center}
\begin{tabular}{ccc}
\hline\hline
 Quantum channel & Useful quantity & Analytical value  \\
\hline
\multirow{3}{5em}{Depolarizing (Sec.~\ref{sec:00000000005B1})} & ${r^D_t}$ & $(1 - {\lambda_t})r$ \\
& ${\nu^D_t}$ & $(1 - \frac{\lambda_t}{2})r$ \\
& ${\sum_{l = 0}^3}\, {\|{K_l}{\rho_0}(d{K_l^{\dagger}}/dt)\|_1}$ & $\frac{3\Gamma}{4}(1 - {\lambda_t})$ \\
\hline
\multirow{3}{7em}{\begin{center}Phase damping (Sec.~\ref{sec:00000000005B2})\end{center}} & ${r^{PD}_t}$ & $r\sqrt{1 - {\lambda_t}\,{\sin^2}\theta}$ \\
& ${\nu^{PD}_t}$ & $r\sqrt{1 - \frac{1}{4}\left({\lambda_t} + 2\left(1 - \sqrt{1 - {\lambda_t}}\,\right)\right){\sin^2}\theta}$ \\
& ${\sum_{l = 0}^1}\, {\|{K_l}{\rho_0}(d{K}_l^{\dagger}/dt)\|_1}$ & $\frac{\Gamma}{4}\left(1 - {\lambda_t}\right)\left(1 - r\cos\theta + \sqrt{{\left(1-r\cos\theta\right)^2} + \frac{{r^2}\,{\sin^2}\theta}{1 - {\lambda_t}}}\,\right)$ \\
\hline
\multirow{3}{7em}{\begin{center}Generalized amplitude damping (Sec.~\ref{sec:00000000005B3})\end{center}} & ${r^{GAD}_t}$ & $\sqrt{{\left[(2\alpha - 1){\lambda_t} + (1 - {\lambda_t})r\cos\theta\right]^2} + (1 - {\lambda_t}){r^2}{\sin^2}\theta}$ \\
& ${\nu^{GAD}_t}$ & $\begin{aligned}&\frac{1}{2}\left\{  {\lambda_t}(2\alpha - 1)\left(2(2 - {\lambda_t})r\cos\theta + {\lambda_t}(2\alpha - 1)\right) + {r^2}\left[{(2 - {\lambda_t})^2} \right.\right.\\ &\left.\left. + \left(2 - (2 - {\lambda_t})\sqrt{1 - {\lambda_t}}\,\right)\sqrt{1 - {\lambda_t}}\,{\sin^2}\theta\right]\right\}^{1/2}\end{aligned}$ \\
& ${\sum_{l = 0}^3}\, {\|{K_l}{\rho_0}(d{K_l^{\dagger}}/dt)\|_1}$ & $\frac{\Gamma\left(1 - {\lambda_t}\right)}{4} {\sum_{s = 1}^2}\, {c_{s\alpha}} \left(1 + {(-1)^s}r\cos\theta + \sqrt{{\left(1 + {(-1)^s}r\cos\theta\right)^2} + \frac{{r^2}{\sin^2}\theta}{1 - {\lambda_t}}} \,\right)$ \\
\hline\hline
\end{tabular}
\label{tab:TABLE01}
\end{center}
\end{table*}

To conclude, we investigate the QSL time based on the QJSD metric. In detail, we find that ${D_{JS}}({\rho_0},{\rho^C_{\tau}}) = \sqrt{S({\varpi^C_{\tau}}) - S({\rho_0})}$, which comes from the fact the von Neumann entropy remains invariant for input states undergoing unitary evolutions, i.e., $S({U_t}{\rho_0}{U_t^{\dagger}}) = S({\rho_0})$ for all $t > 0$. In this case, Eq.~\eqref{eq:0000015} becomes
\begin{equation}
\label{eq:0000022}
{\mathcal{\tau}_{JS}^{\text{QSL}}} := \frac{h(\frac{1 - {r}}{2}) - h(\frac{1 - {\nu^C_{\tau}}}{2})}{r \|\vec{n}\| \|\hat{n}\times\hat{r} \, \| \, {\left\langle\left\langle \left|\ln\left(\frac{(1 - r)(1 - {\nu^C_t})}{4}\right)\right|\right\rangle\right\rangle_{\tau}}} ~,
\end{equation}
where we define 
\begin{equation}
\label{eq:0000023}
h(x) := x\ln{x} + (1 - x)\ln(1 - x) ~,
\end{equation}
with $0 \leq x \leq 1$. We note that Eq.~\eqref{eq:0000022} is fully characterized by the parameters $r$, $\|\vec{n}\|$, and $\tau$, also depending on the Euclidean norm $\|\hat{n}\times\hat{r}\|$ that displays the role of the coherences of the probe state $\rho_0$ with respect to the eigenstates of the Hamiltonian $H$. On the one hand, for $\|\vec{n}\|\tau \ll 1$, Eq.~\eqref{eq:0000022} approaches the value 
\begin{equation}
\label{eq:0000024}
{\mathcal{\tau}_{JS}^{\text{QSL}}} \approx \frac{\|\hat{n}\times\hat{r}\| \, \|\vec{n}\|{\tau^2}}{8\ln[(1 - r)/2]} \ln\left(\frac{1 - r}{1 + r}\right) 
\end{equation} 
at earlier times of the dynamics. On the other hand, for all $t > 0$, using the L'H\^{o}pital rule, we note that ${\lim_{\|\hat{n}\times\hat{r}\| \rightarrow 0}} \, {\mathcal{\tau}_{JS}^{\text{QSL}}} = 0$, i.e., the entropic QSL time is zero-valued for initial incoherent states such that $\|\hat{n}\times\hat{r} \| = 0$, as expected. In particular, for $0 < \|\hat{n}\times\hat{r}\| \ll 1$ and $t > 0$, it can be ve\-ri\-fied that the lowest approximation order of Eq.~\eqref{eq:0000022} with respect to such a quantity is given by 
\begin{equation}
\label{eq:0000025}
{\mathcal{\tau}_{JS}^{\text{QSL}}} \approx \frac{\|\hat{n}\times\hat{r}\| \, {\sin^2}(\|\vec{n}\|\tau)}{8\|\vec{n}\|\ln[(1 - r)/2]} \ln\left(\frac{1 - r}{1 + r}\right) ~,
\end{equation}
which in turn exhibits a harmonic behavior as a function of time.


\subsection{Nonunitary evolution}
\label{sec:00000000005B}

In the following, we address the QSLs related to nonunitary evolutions described by the quantum noisy operations: (i) depola\-ri\-zing channel [see Sec.~\ref{sec:00000000005B1}], (ii) phase damping channel [see Sec.~\ref{sec:00000000005B2}], and (iii) ge\-ne\-ra\-lized amplitude damping channel (GAD) [see Sec.~\ref{sec:00000000005B3}]. The study of these QSLs takes into account Eqs.~\eqref{eq:0000016} and~\eqref{eq:0000017}. Throughout the text, we report analytical expressions for QJPD, QJSD and other quantities useful for calculating such entropic QSLs for each of the quantum channels considered. In Table~\ref{tab:TABLE01}, we summarize the results and provide some additional details.

To investigate the tightness of the QSLs, we consider the figure of merit defined as follows
\begin{equation}
\label{eq:0000026}
{\delta_{J,JS}}(\tau) := 1 - \frac{{\mathcal{\tau}_{J,JS}^{\text{QSL}}}}{\tau} ~.
\end{equation}
In Eq.~\eqref{eq:0000026}, the smaller the relative error, the tighter the entropic speed limit related to QJSD and QJPD. The fi\-gu\-re of merit signals the deviation of the QSL bound concerning the average quantum speed. The bounds saturate whenever ${D^2_{J,JS}}({\rho_0},{\rho_{\tau}})$ coincides with the time-average of the product between the cost function $ {f_{J,JS}}({\rho_0},{\rho_t})$ and the Schatten speed that is induced by the overall dynamics. For convenience, throughout the paper we focus on the normalized relative error ${\widetilde{\delta}_{J,JS}}({\tau})$, with $\widetilde{x} := (x - \text{min}(x))/(\text{max}(x) - \text{min}(x))$, observing that $0 \leq {\widetilde{\delta}_{J,JS}}({\tau}) \leq 1$. To be clear, note that $\text{min}({\delta_{J,JS}}(\tau))$ and $\text{max}({\delta_{J,JS}}(\tau))$ are evaluated from the set of possible values obtained for the relative error over the time interval considered in each of the numerical simulations.


\subsubsection{Depolarizing channel}
\label{sec:00000000005B1}

\begin{figure*}[!t]
\begin{center}
\includegraphics[scale=0.85]{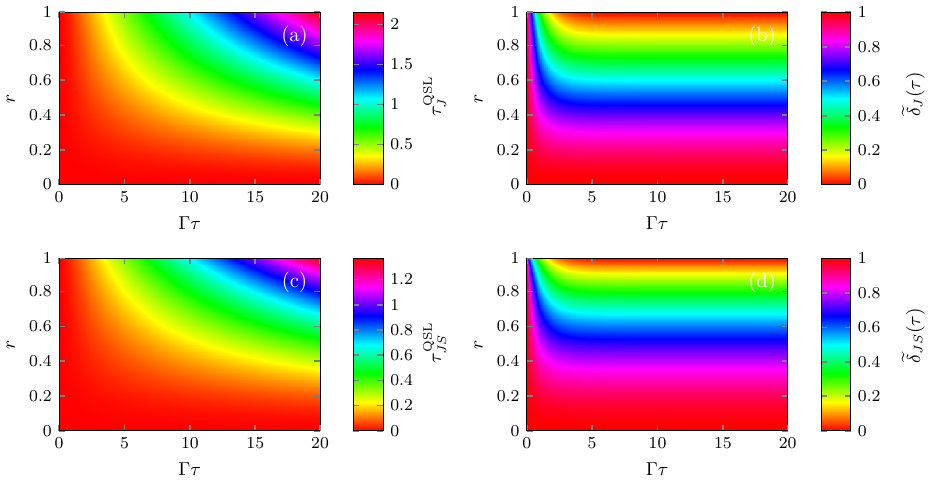}
\caption{(Color online) Density plot of the entropic QSLs ${\tau}_{J,JS}^{\text{QSL}}$ [see Eqs.~\eqref{eq:0000030} and~\eqref{eq:0000036}, respectively], and the normalized relative errors $\widetilde{\delta}_{J,JS}({\tau})$ [see Eq.~\eqref{eq:0000026}], as a function of the dimensionless parameters $\Gamma\tau$ and $r \in [0,1)$, for initial single-qubit states evolving under the depo\-la\-ri\-zing channel [see Sec.~\ref{sec:00000000005B1}]. Here, we consider $\theta \in [0,\pi]$ and $\phi \in [0,2\pi)$.}
\label{fig:FIG00002}
\end{center}
\end{figure*}

The depolarizing channel is described by the Kraus operators ${K_0} = \sqrt{1 - (3{\lambda_t}/4)}\,\mathbb{I}$, and ${K_{1,2,3}} = \sqrt{{\lambda_t}/4}\,{\sigma_{x,y,z}}$, with ${\lambda_t} = 1 - {e^{-\Gamma t}}$, where $\Gamma$ is the damping constant~\cite{NIELSEN}. We note that this quantum channel is unital, i.e, ${\sum_{l = 0}^3}\,{K_l^{\dagger}}{K_l} = {\sum_{l = 0}^3}\,{K_l}{K_l^{\dagger}} =\mathbb{I}$. The evolved state of the two-level system becomes ${\rho^D_t} = {\sum_{l = 0}^3}\,{K_l}{\rho_0}{K_l^{\dagger}} = (1/2)(\mathbb{I} + {r^D_t}(\hat{r}\cdot\vec{\sigma}))$, where we define ${r^D_t} = (1 - {\lambda_t}){r}$. This state has eigenvalues $\kappa_{\text{max}/\text{min}}({\rho^D_t}) = (1 \pm {r^D_t})/2$. We recall that the quantum channel uniformly shrinks the Bloch sphere toward its center, thus fully depola\-ri\-zing the initial single-qubit state. In fact, for $\Gamma{t} \gg 1$, the stationary state ${\rho^{D}_{\infty}} \approx {\mathbb{I}}/2$ is the maximally mixed one. In turn, one finds the state ${\varpi^D_t} = ({\rho_0} + {\rho^D_t})/2 = (1/2)(\mathbb{I} + {\nu^D_t}(\hat{r}\cdot\vec{\sigma}))$, where ${\nu^D_t} = (1 - {\lambda_t}/2)r$, with eigen\-va\-lues $\kappa_{\text{max}/\text{min}}({\varpi^D_t}) = (1/2)(1 \pm {\nu^D_t})$.

In this setting, one finds that the quantum Jeffreys pseudodistance related to initial $\rho_0$ and evolved $\rho^D_{\tau}$ states yields
\begin{align}
\label{eq:0000027}
{D_J}({\rho_0},{\rho^D_{\tau}}) = \frac{1}{2}\sqrt{\left(r - {r_{\tau}^D}\right)\ln\left[\frac{(1 + r)(1 - {r^D_{\tau}})}{(1 - r)(1 + {r^D_{\tau}})}\right]} ~.
\end{align}
Equation~\eqref{eq:0000027} shows that the QJPD is independent of the parameters ${\theta}$ and ${\phi}$, being function of the dimensionless parameters $\Gamma\tau$ and $r$. On the one hand, for ${\Gamma}{\tau} \ll 1$, the QJPD becomes 
\begin{equation}
\label{eq:0000028}
{D_J^2}({\rho_0},{\rho^D_{\tau}}) \approx \frac{{r^2}{(\Gamma\tau)^2}}{2(1 - {r^2})} ~. 
\end{equation}
On the other hand, for $\Gamma \tau \gg 1$, the QJPD approaches the asymptotic value 
\begin{equation}
\label{eq:0000029}
{D_J^2}({\rho_0},{\rho^D_{\tau}}) \approx \frac{r}{4}\ln\left(\frac{1 + r}{1 - r}\right) ~, 
\end{equation}
being fully characterized by the mixedness parameter $r \in [0,1)$. The latter result is in agreement with the fact that the system approaches the maximally mixed state $\mathbb{I}/2$ for ${\Gamma}{\tau} \gg 1$.

Next, the QSL time based on the QJPD is evaluated as follows
\begin{align}
\label{eq:0000030}
&\frac{\mathcal{\tau}_{J}^{\text{QSL}}}{\tau} = \frac{r}{3}(1 - {e^{-\Gamma \tau}})\left[\ln\left(\frac{1 + r}{1 - r}\right) - \ln\left(\frac{1 + {r^D_{\tau}}}{1 - {r^D_{\tau}}}\right)\right] \nonumber\\
&\times\left[\left(1 + {e^{-\Gamma\tau}}\right) \ln\left(1 - r{e^{-\Gamma\tau}}\right) - \left(3 - {e^{-\Gamma\tau}}\right)\ln(1 - r) \right. \nonumber\\
&\left. + \left(1 - {e^{-\Gamma\tau}}\right)(2\ln2 + 1)\right]^{-1} ~.
\end{align}
On the one hand, for ${\Gamma}{\tau} \ll 1$, we verify that Eq.~\eqref{eq:0000030} approaches as
\begin{equation}
\label{eq:0000031}
\frac{\mathcal{\tau}_J^{\text{QSL}}}{\tau} \approx \frac{2{r^2}{{\Gamma}{\tau}}}{3(1 + r)[1 + r + (1 - r)(\ln4 - 2\ln(1 - r))]} ~,
\end{equation}
which increases linearly with ${\Gamma}{\tau}$ at earlier times of the dy\-na\-mics. On the other hand, for ${\Gamma}{\tau} \gg 1$, one obtains the ratio 
\begin{equation}
\label{eq:0000032}
\frac{{\mathcal{\tau}_J^{\text{QSL}}}}{\tau} \approx \frac{r\ln[(1 + r)/(1 - r)]}{3[2\ln2 + 1 - 3\ln(1 - r)]} ~, 
\end{equation}
which in turn stands as a time-independent quantity at later times of the dy\-na\-mics.

In turn, the quantum Jensen-Shannon distance can be written as
\begin{align}
\label{eq:0000033}
&{D_{JS}}({\rho_0},{\rho^D_{\tau}}) = \nonumber\\
& \frac{1}{\sqrt{2}}\sqrt{h\left(\frac{1 - r}{2}\right) + h\left(\frac{1 - {r^D_\tau}}{2}\right)- 2h\left(\frac{1 - {\nu^D_\tau}}{2}\right)} ~,
\end{align}
with $h(\bullet)$ defined in Eq.~\eqref{eq:0000023}. It is worth noting that QJSD is a function of $r$ and $\Gamma\tau$. In this case, the mixedness degree of the initial state $\rho_0$ strongly accounts for its distinguishability from any instantaneous single-qubit state. In particular, for $\Gamma\tau \ll 1$, one finds that the squared QJSD becomes 
\begin{equation}
\label{eq:0000034}
{D_{JS}^2}({\rho_0},{\rho^D_{\tau}}) \approx \frac{{r^2}{(\Gamma\tau)^2}}{8(1 - {r^2})} ~,
\end{equation}
thus increasing quadratically with the dimensionless parameter $\Gamma\tau$. However, for $\Gamma\tau \gg 1$, the squared QJSD approaches the asymptotic value
\begin{equation}
\label{eq:0000035}
{D_{JS}^2}({\rho_0},{\rho^D_{\tau}}) \approx \frac{1}{2}\left[h\left(\frac{1 - r}{2}\right) - \ln{2}\right] - h\left(\frac{2 - r}{4}\right) ~,
\end{equation}
which is a time-independent quantity.

\begin{figure*}[!t]
\begin{center}
\includegraphics[scale=0.85]{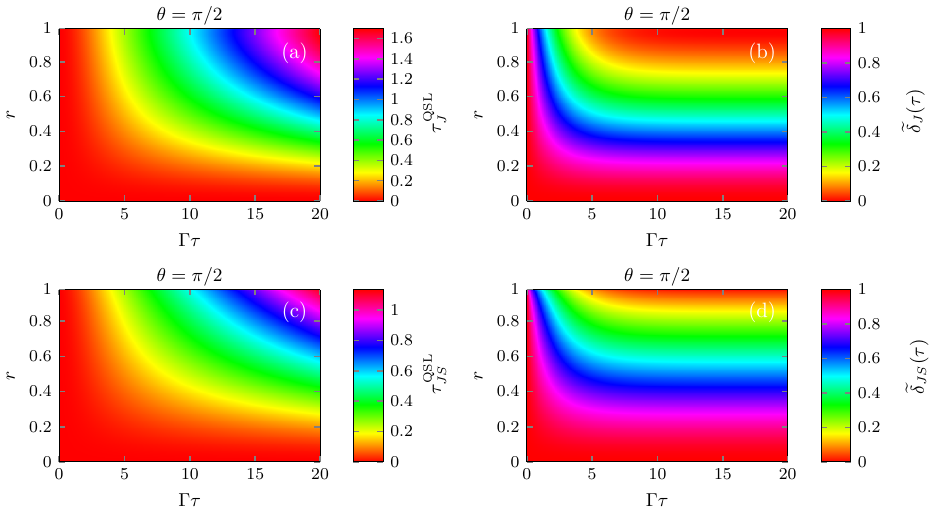}
\caption{(Color online) Density plot of the entropic QSLs ${\tau}_{J,JS}^{\text{QSL}}$ [see Eqs.~\eqref{eq:0000016} and~\eqref{eq:0000017}, respectively], and the normalized relative errors $\widetilde{\delta}_{J,JS}({\tau})$ [see Eq.~\eqref{eq:0000026}], as a function of the dimensionless parameters $\Gamma\tau$ and $r \in [0,1)$, for initial single-qubit states subject to the phase damping quantum channel [see Sec.~\ref{sec:00000000005B2}]. Here, we set $\theta = \pi/2$, for all $\phi \in [0,2\pi)$.}
\label{fig:FIG00003}
\end{center}
\end{figure*}

In addition, the QSL time in Eq.~\eqref{eq:0000017} related to the quantum Jensen-Shannon distance becomes
\begin{align}
\label{eq:0000036}
&\frac{\mathcal{\tau}_{JS}^{\text{QSL}}}{\tau} = \frac{2r}{3}\left[h\left(\frac{1 - r}{2}\right) + h\left(\frac{1 - {r^D_{\tau}}}{2}\right) - 2 h\left(\frac{1 - {\nu^D_{\tau}}}{2}\right)\right]\nonumber\\
&\times \left\{ (2 + \ln{2})\left(r - {r^D_{\tau}}\right) - \left(1 - {r^D_{\tau}}\right)\ln\left[({1 - {r^D_{\tau}}})/{4}\right] \right.\nonumber\\
&\left. - \left(2 - r - {r^D_{\tau}}\right)\ln\left(2 - r - {r^D_{\tau}}\right) + 3(1 - r)\ln(1 - r)\right\}^{-1} ~.
\end{align}
From Eq.~\eqref{eq:0000036}, we find that the ratio ${{\tau}_{JS}^{\text{QSL}}}/\tau$ is independent of the angular coordinates ${\theta}$ and ${\phi}$ of the Bloch sphere. We find the ratio ${\mathcal{\tau}_{JS}^{\text{QSL}}}/{\tau} \rightarrow 0$ for $\Gamma\tau \rightarrow 0$, while for $\Gamma\tau \gg 1$ we have that the QSL time approaches a time-independent quantity that exclusively depends on the mixedness parameter $r \in [0,1)$. It turns out that ${{\tau}_{JS}^{\text{QSL}}}/\tau \rightarrow 0$ for $r \rightarrow 0$, regardless of whether $\Gamma\tau \geq 0$.

In Fig.~\ref{fig:FIG00002}, we plot the speed limits ${\tau}_{J,JS}^{\text{QSL}}$ [see Figs.~\ref{fig:FIG00002}(a) and~\ref{fig:FIG00002}(c)] and the normalized relative errors $\widetilde{\delta}_{J,JS}({\tau})$ [see Figs.~\ref{fig:FIG00002}(b) and~\ref{fig:FIG00002}(d)], as a function of the mixedness parameter $r$ and the dimensionless parameter ${\Gamma}{\tau}$, for initial single-qubit states evolving under the depolarizing channel. Figures~\ref{fig:FIG00002}(a) and~\ref{fig:FIG00002}(c) show that, respectively, ${\tau_J^{\text{QSL}}}$ [see Eq.~\eqref{eq:0000030}] and ${\tau_{JS}^{\text{QSL}}}$ [see Eq.~\eqref{eq:0000036}] exhibit similar qua\-li\-ta\-ti\-ve behavior, thus being monotonic functions of the parameters $r$ and $\Gamma\tau$. In both cases, for a fixed value $r \in [0,1)$ ($\Gamma\tau \geq 0$), the QSL time ${\tau}_{J,JS}^{\text{QSL}}$ increases monotonically as a function of $\Gamma\tau \geq 0$ ($0 \leq r < 1$). In particular, for $0 \leq r \lesssim 0.2$, one finds that ${\tau}_{J,JS}^{\text{QSL}}$ approaches small values for the overall dynamics. This means that the closer the initial state is to the center of the Bloch sphere, the faster the system evolves toward to the ma\-xi\-mally mixed state.

Figures~\ref{fig:FIG00002}(b) and~\ref{fig:FIG00002}(d) show that $\widetilde{\delta}_{J,JS}({\tau})$ presents si\-mi\-lar qualitative behavior for both the QJPD and QJSD measures. We note that, for a fixed value $\Gamma\tau \geq 0$, then $\widetilde{\delta}_{J,JS}({\tau})$ monotonically decreases as a function of $r$. In particular, for $0.9 \lesssim r < 1$ and $\Gamma\tau \gtrsim 3$, one finds that the normalized relative errors approaches small values $\widetilde{\delta}_{J,JS}({\tau}) \approx 0$. This suggests that the closer the initial state is to a pure state, that is, the further it is from the fixed point of the nonunitary dynamics, the tighter the bounds will be. Otherwise, the bounds are looser for initial states close to the maximally mixed state, regardless of whether $\Gamma\tau \geq 0$. We note that, for a given fixed value $0 \leq r < 1$, each relative error approaches a certain value that remains constant at later times of the dynamics.


\subsubsection{Phase damping channel}
\label{sec:00000000005B2}

The phase damping channel is described by the Kraus operators ${K_0} = |0\rangle\langle{0}| + \sqrt{1 - {\lambda_t}}|1\rangle\langle{1}|$, and ${K_1} = \sqrt{{\lambda_t}}|1\rangle\langle{1}|$, where $\left\{|0\rangle,|1\rangle\right\}$ is the computational basis states, with ${\lambda_t} = 1 - {e^{-\Gamma t}}$, and $\Gamma$ is the deco\-he\-ren\-ce rate~\cite{NIELSEN}. This is also a unital quantum channel, which in turn describes the loss of coherence of open quantum systems without energy dissipation. In this case, the energy spectrum of the system remains unchanged, while its eigenstates accumulate a relative phase that is destroyed during the nonunitary evolution.

\begin{figure*}[!t]
\begin{center}
\includegraphics[scale=0.825]{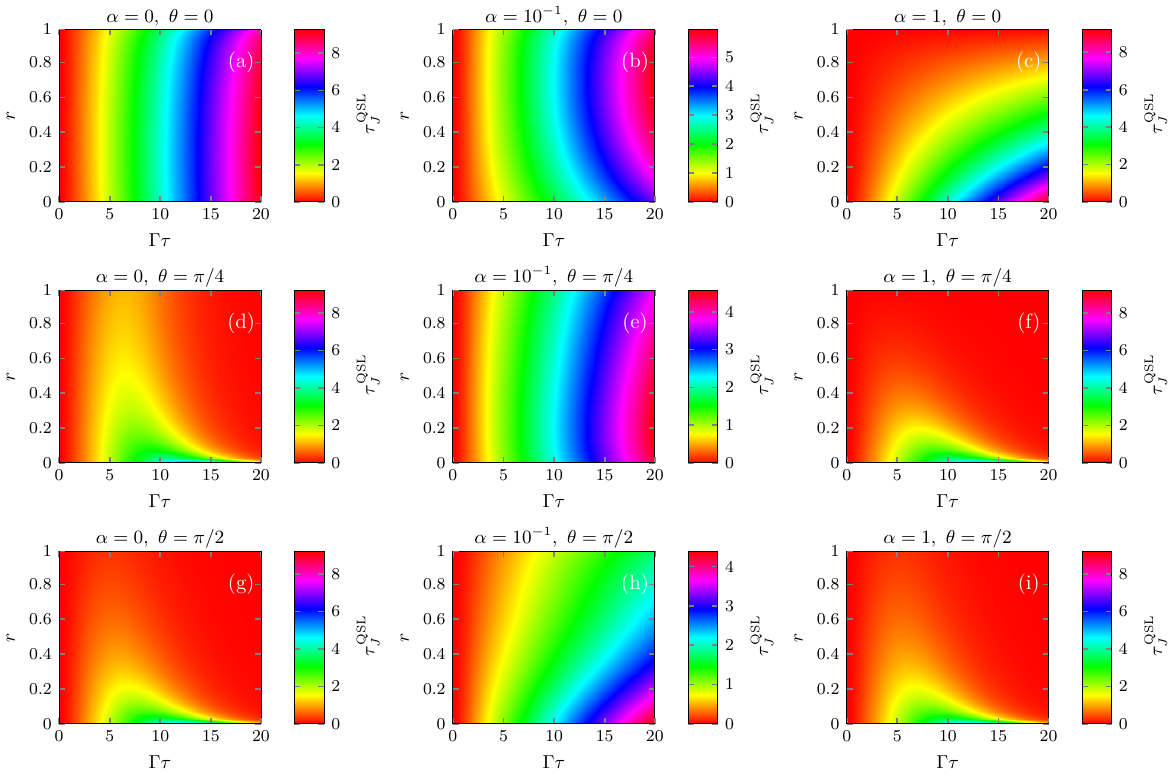}
\caption{(Color online) Density plot of the entropic QSL ${\tau}_J^{\text{QSL}}$ [see Eq.~\eqref{eq:0000016}], as a function of the dimensionless parameters $\Gamma\tau$ and $r \in [0,1)$, for initial single-qubit states subject to the generalized amplitude damping channel [see Sec.~\ref{sec:00000000005B3}]. Here, we set $\alpha = \{0,0.1,1\}$ and $\theta = \{0,\pi/4,\pi/2$\}, for all $\phi \in [0,2\pi)$.}
\label{fig:FIG00004}
\end{center}
\end{figure*}

The evolved state of the two-level system is given by ${\rho^{PD}_t} = {\sum_{j = 0}^1}\,{K_j}{\rho_0}{K_j^{\dagger}} = (1/2)\left(\mathbb{I} + {\vec{r}^{\,PD}_t}\cdot\vec{\sigma}\right)$, with ${\vec{r}^{\,PD}_t} = {\eta_t^{PD}}\vec{r}$, where ${\eta_t^{PD}} = \text{diag}(\sqrt{1 - {\lambda_t}},\sqrt{1 - {\lambda_t}},1)$ is a real-valued, diagonal matrix. For $\Gamma{t} \gg 1$, one obtains the asymptotic vector ${\vec{r}_{\infty}^{\,PD}} \approx (0,0,r\cos\theta)$ related to the stationary state ${\rho^{PD}_{\infty}} \approx (1/2)((1 + r\cos\theta)|0\rangle\langle{0}| + (1 - r\cos\theta)|1\rangle\langle{1}|)$. The latter is an incoherent state with respect to the computational basis $\{|0\rangle,|1\rangle\}$, i.e., the complete set of eigenvectors of the observable $\sigma_z$. The instantaneous state exhi\-bits the eigenvalues ${\kappa_{\text{max}/\text{min}}}({\rho^{PD}_t}) = (1 \pm {r^{PD}_t})/2$, with 
\begin{equation}
\label{eq:0000037}
{r^{PD}_t} = r\sqrt{1 - {\lambda_t}\,{\sin^2}\theta} ~. 
\end{equation}
In turn, for the convex combination ${\varpi^{PD}_t} = (1/2)({\rho_0} + {\rho^{PD}_t})$, we find the time-dependent eigen\-va\-lues ${\kappa_{\text{max}/\text{min}}}({\varpi^{PD}_t}) = (1/2)(1 \pm {\nu^{PD}_t})$, with 
\begin{equation}
\label{eq:0000038}
{\nu^{PD}_t} = r\sqrt{1 - \frac{1}{4}\left({\lambda_t} + 2\left(1 - \sqrt{1 - {\lambda_t}}\,\right)\right){\sin^2}\theta} ~.
\end{equation}

In this setting, we find the analytical expression for the quantum Jeffreys pseudodistance given as follows
\begin{align}
\label{eq:0000039}
&{D_J}({\rho_0},{\rho^{PD}_{\tau}}) = \frac{\sin\theta}{2}\sqrt{r\left(1 - \sqrt{1 - {\lambda_{\tau}}}\right)}\left[\ln\left(\frac{1 + r}{1 - r}\right) \right. \nonumber\\
 &\left. -\frac{r}{r^{PD}_{\tau}}\sqrt{1 - {\lambda_{\tau}}}\ln\left(\frac{1 + {r^{PD}_{\tau}}}{1 - {r^{PD}_{\tau}}}\right)\right]^{1/2} ~.
\end{align}
Equation~\eqref{eq:0000039} shows that the quantum Jeffreys pseudodistance does not depend on the initial azimuth angle ${\phi}$. We note that the QJPD vanishes for incoherent probe states with $\theta = 0$ that belong to the $z$ axis of the Bloch sphere. On the one hand, for $\Gamma\tau \ll 1$, we find that the QJPD becomes 
\begin{align}
\label{eq:0000040}
&{D_J^2}({\rho_0},{\rho^{PD}_{\tau}}) \approx \nonumber\\
&\frac{r}{32}{(\Gamma\tau)^2}{\sin^2}(2\theta)\left[\frac{1}{2}\ln\left(\frac{1 + r}{1 - r}\right) + \frac{r{\tan^2}\theta}{(1 - {r^2})}\right] ~, 
\end{align}
which in turn scales quadratically with the dimensionless parameter $\Gamma\tau$. On the other hand, for $\Gamma\tau \gg 1$, the QJPD approaches the asymptotic value 
\begin{equation}
\label{eq:0000041}
{D_J^2}({\rho_0},{\rho^{PD}_{\tau}}) \approx \frac{r}{4} \, {\sin^2}{\theta}\ln\left(\frac{1 + r}{1 - r}\right) ~, 
\end{equation}
which is a time-independent quantity.

\begin{figure*}[!t]
\begin{center}
\includegraphics[scale=0.825]{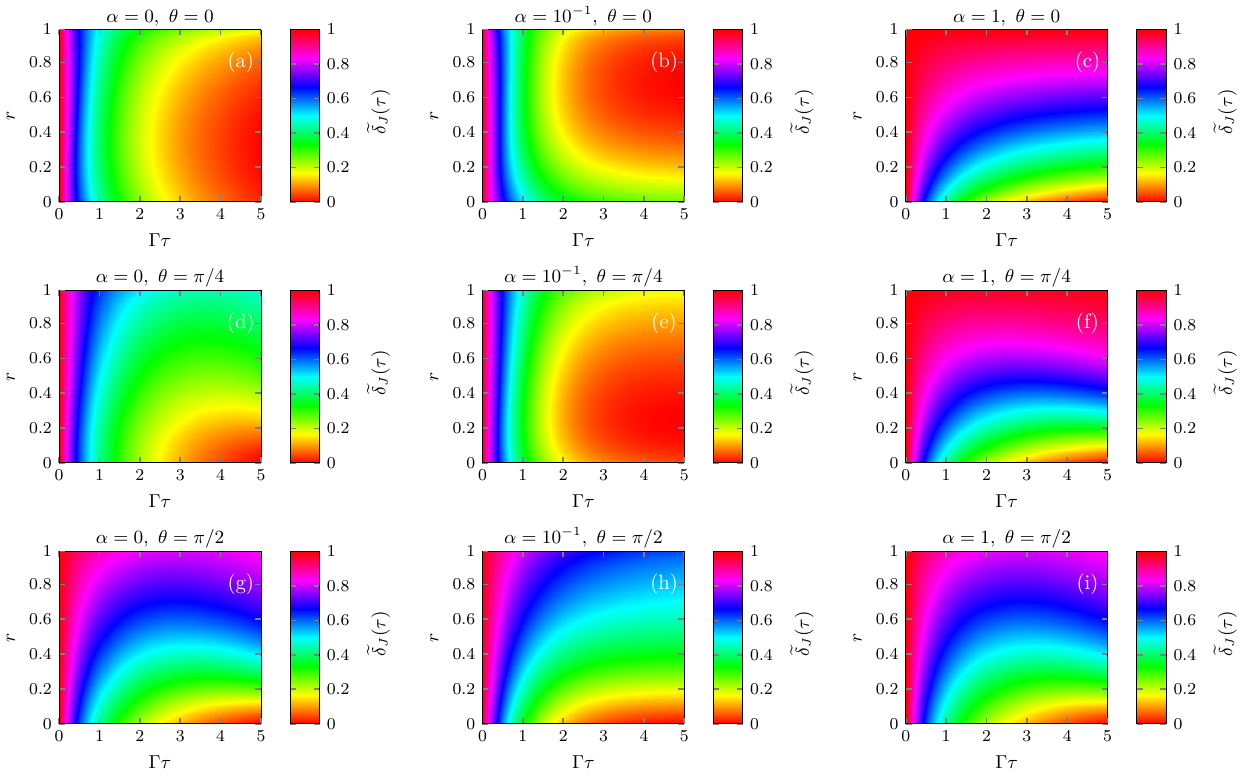}
\caption{(Color online) Density plot of the normalized relative error ${\widetilde{\delta}_J}({\tau})$ [see Eq.~\eqref{eq:0000026}], as a function of the dimensionless parameters $\Gamma\tau$ and $r \in [0,1)$, for initial single-qubit states subject to the generalized amplitude damping channel [see Sec.~\ref{sec:00000000005B3}]. Here, we set $\alpha = \{0,0.1,1\}$ and $\theta = \{0,\pi/4,\pi/2$\}, for all $\phi \in [0,2\pi)$.}
\label{fig:FIG00005}
\end{center}
\end{figure*}

In turn, one finds that the quantum Jensen-Shannon distance is given by
\begin{align}
\label{eq:0000042}
&{D_{JS}}({\rho_0},{\rho^{PD}_{\tau}}) = \nonumber\\
& \frac{1}{\sqrt{2}}\sqrt{h\left(\frac{1 - {r^{PD}_{\tau}}}{2}\right) + h\left(\frac{1 - r}{2}\right) - 2h\left(\frac{1 - {\nu^{PD}_{\tau}}}{2}\right)} ~.
\end{align}
On the one hand, for $\Gamma\tau \ll 1$, one verifies that Eq.~\eqref{eq:0000042} can be written as
\begin{align}
\label{eq:0000043}
&{D^2_{JS}}({\rho_0},{\rho^{PD}_{\tau}}) \approx \nonumber\\
& \frac{r}{128}{(\Gamma\tau)^2}{\sin^2}(2\theta)\left[\frac{1}{2}\ln\left(\frac{1 + r}{1 - r}\right) + \frac{r{\tan^2}\theta}{(1 - {r^2})}\right] ~, 
\end{align}
which means that the squared QJSD increases quadratically with the dimensionless parameter $\Gamma\tau$. On the other hand, for $\Gamma\tau \gg 1$, the QJSD approaches the time-independent value 
\begin{align}
\label{eq:0000044}
&{D_{JS}^2}({\rho_0},{\rho^{PD}_{\tau}}) \approx \frac{1}{2}\left[h\left(\frac{1 - r\cos\theta}{2}\right) + h\left(\frac{1 - r}{2}\right) \right.\nonumber\\
&\left.- 2h\left(\frac{2 - r\sqrt{4 - 3{\sin^2}\theta}}{4}\right)\right] ~.
\end{align}

\begin{figure*}[!t]
\begin{center}
\includegraphics[scale=0.825]{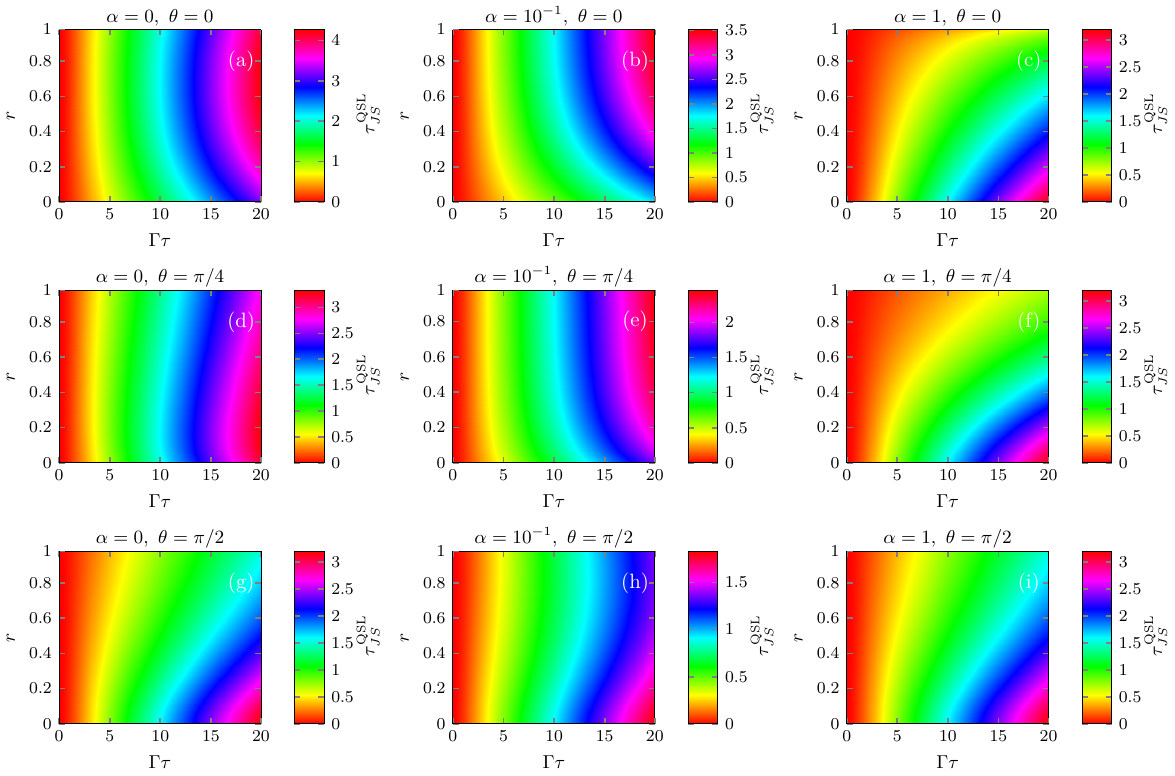}
\caption{(Color online) Density plot of the entropic QSL ${\tau}_{JS}^{\text{QSL}}$ [see Eq.~\eqref{eq:0000017}], as a function of the dimensionless parameters $\Gamma\tau$ and $r \in [0,1)$, for initial single-qubit states subject to the generalized amplitude damping channel [see Sec.~\ref{sec:00000000005B3}]. Here, we set $\alpha = \{0,0.1,1\}$ and $\theta = \{0,\pi/4,\pi/2$\}, for all $\phi \in [0,2\pi)$.}
\label{fig:FIG00006}
\end{center}
\end{figure*}

The entropic QSL times ${\tau_{J,JS}^{\text{QSL}}}$ related to the QJPD and QJSD are evaluated through Eqs.~\eqref{eq:0000016} and~\eqref{eq:0000017}, respectively. The expressions are too long to be reported here. We note that the cost functions ${f_{J,JS}}({\rho_0},{{\rho^{PD}_t}})$ are given by Eqs.~\eqref{eq:0000008} and~\eqref{eq:0000009}, respectively. In Table~\ref{tab:TABLE01}, we show useful information-theoretic parameters needed to compute ${\tau_{J,JS}^{\text{QSL}}}$ for initial single-qubit states evolving under the phase damping quantum channel. Overall, one arrives at trivial bounds ${\tau_{J,JS}^{\text{QSL}}} \approx 0$ for initial states belonging to the $z$ axis of the Bloch sphere, since they are invariant under the quantum operation.

In Fig.~\ref{fig:FIG00003}, we show the density plots of the entropic QSL time $\tau^{\text{QSL}}_{J,JS}$ [see Figs.~\ref{fig:FIG00003}(a) and~\ref{fig:FIG00003}(c)], and the normalized relative error ${\widetilde{\delta}_{J,JS}}(\tau)$ [see Figs.~\ref{fig:FIG00003}(b) and~\ref{fig:FIG00003}(d)], as a function of the dimensionless parameters $\Gamma\tau$ and $r$, for single-qubit states evolving under the phase damping channel. Here we set initial states with $\theta = \pi/2$ lying in the equatorial plane of the Bloch sphere. Figures~\ref{fig:FIG00003}(a) and~\ref{fig:FIG00003}(c) show that $\tau^{\text{QSL}}_J$ and $\tau^{\text{QSL}}_{JS}$ exhibit the same qualitative behavior, monotonically increasing as a function of $\Gamma\tau$ and $r$. It is worth noting that ${\tau_{J,JS}^{\text{QSL}}}$ approaches small values at earlier times of the dynamics ($0 \leq \Gamma\tau \lesssim 2$), regardless the purity of the initial state. In addition, for all $\Gamma\tau \geq 0$, the less pure the probe state, the smaller the entropic speed limit time.

In Figs.~\ref{fig:FIG00003}(b) and~\ref{fig:FIG00003}(d), we plot the normalized relative errors ${\widetilde{\delta}_{J,JS}}(\tau)$ as a function of the dimensionless pa\-ra\-me\-ters $\Gamma\tau$ and $r$. On the one hand, for fixed $\Gamma\tau \geq 0$, one finds that ${\widetilde{\delta}_{J,JS}}(\tau)$ decreases as a function of the mixedness parameter $r \in [0,1)$. On the other hand, for fixed $r \gtrsim 0.2$, we have that ${\widetilde{\delta}_{J,JS}}(\tau)$ monotonically decreases as a function of $\Gamma\tau$, and approaches a constant value at later times of the dynamics. The closer the probe state is to the maximally mixed state, the looser the bounds will be, for all $\Gamma\tau \geq 0$. We note that ${\widetilde{\delta}_{J,JS}}(\tau)$ approaches small values at later times of the dynamics for initial states with purity close to the unity.

\begin{figure*}[!t]
\begin{center}
\includegraphics[scale=0.825]{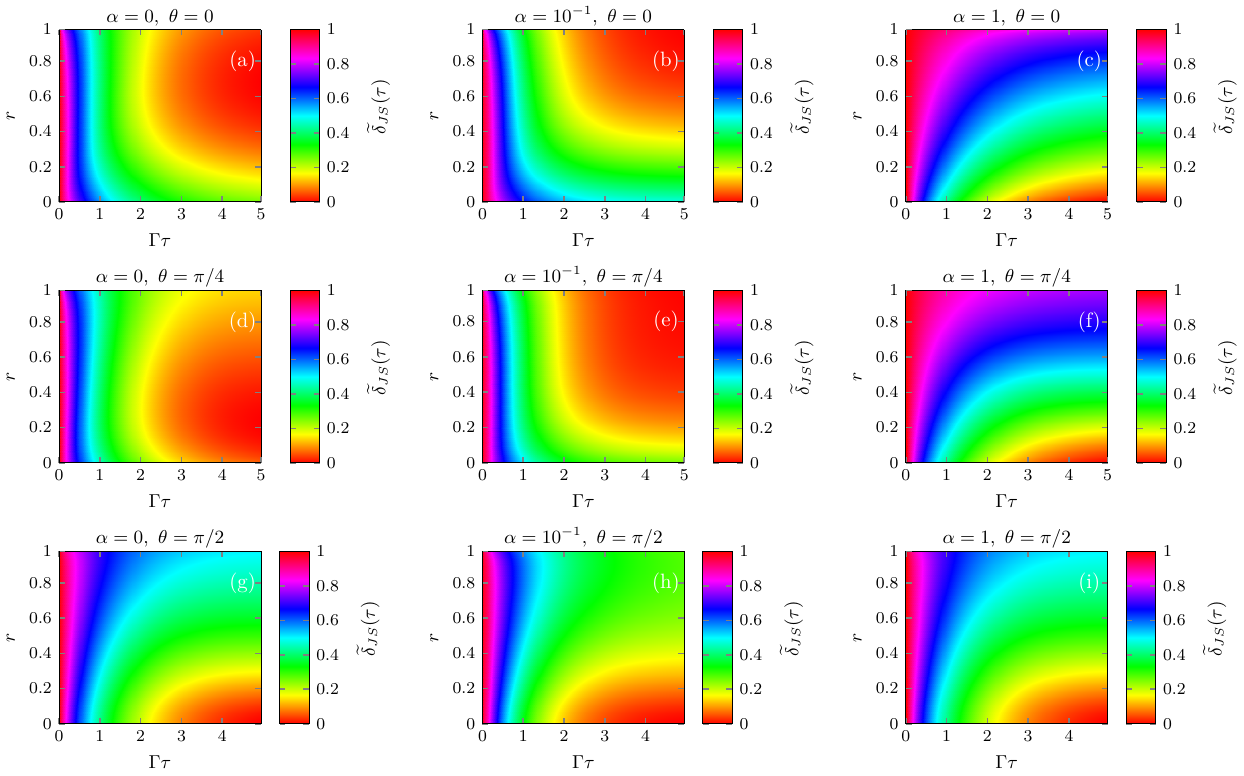}
\caption{(Color online) Density plot of the normalized relative error ${\widetilde{\delta}_{JS}}({\tau})$ [see Eq.~\eqref{eq:0000026}], as a function of the dimensionless parameters $\Gamma\tau$ and $r \in [0,1)$, for initial single-qubit states subject to the generalized amplitude damping channel [see Sec.~\ref{sec:00000000005B3}]. Here, we set $\alpha = \{0,0.1,1\}$ and $\theta = \{0,\pi/4,\pi/2$\}, for all $\phi \in [0,2\pi)$.}
\label{fig:FIG00007}
\end{center}
\end{figure*}


\subsubsection{Generalized amplitude damping channel}
\label{sec:00000000005B3}

The generalized amplitude damping channel (GAD) describes the dissipative dynamics of a two-level system in contact with a thermal bath at finite temperature. The GAD channel models the energy loss from the two-level system to the thermal bath, accounting for excitations between ground and excited states at finite temperature~\cite{GAD001,GAD002}. The quantum channel is modeled by the Kraus operators ${K_0} = \sqrt{\alpha}(|0\rangle\langle{0}| + \sqrt{1 - {\lambda_t}}|1\rangle\langle{1}|)$, ${K_1} = \sqrt{\alpha{\lambda_t}}|0\rangle\langle{1}|$, ${K_2} = \sqrt{1 - \alpha}(\sqrt{1 - {\lambda_t}}|0\rangle\langle{0}| + |1\rangle\langle{1}|)$, and ${K_3} = \sqrt{(1 - \alpha){\lambda_t}}|1\rangle\langle{0}|$, with ${\lambda_t} = 1 - {e^{-\Gamma t}}$, where $1/\Gamma$ is the characteristic time of the dissipative process~\cite{NIELSEN}. The parameter $\alpha \in [0,1]$ encodes the temperature effects of the noisy dynamics, with $\alpha = 1/2$ ($\alpha = 1$) respective to the case of infinite (zero) temperature. In particular, for $\alpha = 1$, GAD channel reduces to the standard amplitude damping channel~\cite{GAD001}.

In this setting, the evolved state becomes ${\rho^{GAD}_t} = {\sum_{j = 0}^3}\,{K_j}{\rho_0}{K_j^{\dagger}} = (1/2)\left(\mathbb{I} + {\vec{r}^{\,GAD}_t}\cdotp\vec{\sigma}\right)$, with ${\vec{r}^{\,GAD}_t} = {\eta_t^{GAD}}\vec{r} + {\vec{\kappa}^{\,GAD}_t}$, where ${\eta_t^{GAD}} = \text{diag}(\sqrt{1 - {\lambda_t}},\sqrt{1 - {\lambda_t}},1 - {\lambda_t})$ is a dia\-go\-nal matrix, while ${\vec{\kappa}^{\,GAD}_t} = (0,0,(2\alpha - 1){\lambda_t})$ is a real-valued tridimensional vector. In particular, for $\Gamma t \gg 1$, one finds the asymptotic vector ${\vec{r}^{\,GAD}_{\infty}} \approx (0,0,2\alpha - 1)$, which in turn describes the stationary, mixed, single-qubit state ${\rho^{GAD}_{\infty}} \approx \alpha |0\rangle\langle{0}| + (1 - \alpha) |1\rangle\langle{1}|$. This means that the GAD channel shrinks the Bloch sphere along the $z$ axis toward to a fixed point located at $z = 2\alpha - 1$~\cite{NIELSEN}. 

The instantaneous state exhibits the eigenvalues ${\kappa_{\text{max}/\text{min}}}({\rho^{GAD}_t}) = (1/2)(1\pm {r^{GAD}_t})$, with
\begin{align}
\label{eq:0000045}
&{r_t^{GAD}} = \nonumber\\
&\sqrt{{\left[(2\alpha - 1){\lambda_t} + (1 - {\lambda_t})r\cos\theta\right]^2} + (1 - {\lambda_t}){r^2}{\sin^2}\theta} ~.
\end{align}
In particular, for $\Gamma{t} \gg 1$ (or $t \rightarrow \infty$), Eq.~\eqref{eq:0000045} approaches the value ${r_{\infty}^{GAD}} \approx |2{\alpha} - 1|$, which in turn depicts the asymptotic state of the Markovian dynamics, while for $t = 0$ one finds that ${r_0^{GAD}} = r$, thus recovering the initial state of the system, as expected. For the convex combination ${\varpi^{GAD}_t} = (1/2)({\rho_0} + {\rho^{GAD}_t})$, we find the set of eigenvalues ${\kappa_{\text{max}/\text{min}}}({\varpi^{GAD}_t}) = (1/2)(1 \pm {\nu^{GAD}_t})$, with
\begin{align}
\label{eq:0000046}
&{\nu^{GAD}_t} = \frac{1}{2}\left\{\lambda_{t}(2\alpha - 1)\left(2(2 - {\lambda_t})r\cos\theta + {\lambda_t}(2\alpha - 1)\right) \right.\nonumber\\
&\left. + {r^2}[(2 - (2 - {\lambda_t})\sqrt{1 - {\lambda_t}}\,)\sqrt{1 - {\lambda_t}}\,{\sin^2}\theta \right.\nonumber\\
&\left. + {(2 - {\lambda_t})^2}]\right\}^{1/2} ~.
\end{align}
In particular, for $t = 0$, Eq.~\eqref{eq:0000046} reduces to ${\nu^{GAD}_0} = r$, which agrees with the fact that ${\varpi_0^{GAD}} = {\rho_0}$ re\-co\-vers the initial state. Furthermore, for $\Gamma{t} \gg 1$, we have that ${\nu^{GAD}_{\infty}} \approx (1/2)\sqrt{{(r\cos\theta + 2\alpha - 1)^2} + {r^2}{\sin^2}\theta}$. This indicates that the states $\rho_0$ and $\rho^{GAD}_{\infty}$ do not commute with each other, and the asymptotic eigenvalues ${\kappa_{\text{max}/\text{min}}}({\varpi^{GAD}_{\infty}})$ still depend on the parameters $(r,\theta,\alpha)$.

Based on these results, it follows that the quantum Jeffreys pseudodistance for the single-qubit states ${\rho_0}$ and ${\rho^{GAD}_{\tau}}$ can be analytically evaluated as follows
\begin{align}
\label{eq:0000047}
&{D_J}({\rho_0},{\rho^{GAD}_{\tau}}) = \frac{1}{2}\left\{{\xi_{\tau}} \ln\left(\frac{1 + r}{1 - r}\right) \right.\nonumber\\
&\left.+ \left({r_{\tau}^{GAD}} + \frac{r({\xi_{\tau}} - r)}{r_{\tau}^{GAD}}\right)\ln\left(\frac{1 + {r^{GAD}_{\tau}}}{1 - {r^{GAD}_{\tau}}}\right)\right\}^{1/2} ~,
\end{align}
where we introduce the auxiliary parameter 
\begin{align}
\label{eq:0000048}
&{\xi_{\tau}} := r\left(1 - \sqrt{1 - {\lambda_{\tau}}}\,\right)\left(1 + \sqrt{1 - {\lambda_{\tau}}}\,{\cos^2}\theta\right) \nonumber\\
&- (2\alpha - 1){\lambda_{\tau}}\cos\theta ~. 
\end{align}
In turn, the quantum Jensen-Shannon distance for ${\rho_0}$ and ${\rho^{GAD}_{\tau}}$ is given by
\begin{align}
\label{eq:0000049}
&{D_{JS}}({\rho_0},{\rho^{GAD}_{\tau}}) = \nonumber\\
& \frac{1}{\sqrt{2}}\sqrt{h\left(\frac{1 - {r^{GAD}_{\tau}}}{2}\right) + h\left(\frac{1 - r}{2}\right) - 2h\left(\frac{1 - {\nu^{GAD}_{\tau}}}{2}\right)} ~.
\end{align}
It is noteworthy that, for all $0 \leq r < 1$ and $\Gamma\tau > 0$, we have that ${D_{J,JS}}({\rho_0},{\rho^{GAD}_{\tau}})$ remains invariant under the mapping $(\theta,\alpha) \rightarrow (\theta \pm \pi, 1 - \alpha)$, with $\theta \in [0,\pi]$ and $0 \leq \alpha \leq 1$. This property is inherited from Eqs.~\eqref{eq:0000045} and~\eqref{eq:0000046}, which means that the eigenvalues ${\kappa_{\text{max}/\text{min}}}({\rho^{GAD}_t})$ and ${\kappa_{\text{max}/\text{min}}}({\varpi^{GAD}_t})$ exhibit such a symmetry.

The entropic QSL times ${\tau_{J,JS}^{\text{QSL}}}$ related to the QJPD and QJSD are evaluated through Eqs.~\eqref{eq:0000016} and~\eqref{eq:0000017}, respectively. The expressions are too long to be reported here. In Table~\ref{tab:TABLE01}, we show some useful quantities needed to compute ${\tau_{J,JS}^{\text{QSL}}}$ for initial single-qubit states evolving under the generalized amplitude damping channel. We recall that the cost functions ${f_{J,JS}}({\rho_0},{{\rho^{GAD}_t}})$ are given in Eqs.~\eqref{eq:0000008} and~\eqref{eq:0000009}, respectively. It turns out that ${\tau_{J,JS}^{\text{QSL}}}$ and ${\widetilde{\delta}_{J,JS}}({\tau})$ are symmetric under the mapping $(\theta,\alpha) \rightarrow (\theta \pm \pi, 1 - \alpha)$, this being an inherited pro\-per\-ty from the QJPD and QJSD. In other words, given two probe states with $(r,{\theta_1},\phi,{\alpha_1})$ and $(r,{\theta_2},\phi,{\alpha_2})$, one expects to obtain the same QSL time whenever we have $|{\theta_{1(2)}} - {\theta_{2(1)}}| = \pi$ and ${\alpha_1} + {\alpha_2} = 1$, for all $\Gamma{\tau} > 0$.

Hereafter, we analyze numerical simulations of the entropic QSL times ${\tau_{J,JS}^{\text{QSL}}}$ and the normalized relative errors ${\widetilde{\delta}_{J,JS}}({\tau})$ for single-qubit states evolving under the GAD channel. We set probe states with $\theta = \{0,\pi/4,\pi/2\}$, regardless the a\-zi\-muth angle $\phi \in [0,2\pi)$, and also consider the parameters $\alpha = \{0,0.1,1\}$. 

In Fig.~\ref{fig:FIG00004}, we show the density plots of the entropic QSL time ${\tau_J^{\text{QSL}}}$ as a function of the dimensionless parameters $\Gamma\tau$ and $r$. We note that, for a fixed value $r \in [0,1)$ of the mixedness parameter, with $\alpha = 0.1$ and $\theta = \{0,\pi/4,\pi/2\}$, then ${\tau_J^{\text{QSL}}}$ monotonically increases as a function of the dimensionless parameter $\Gamma\tau \geq 0$ [see Figs.~\ref{fig:FIG00004}(b),~\ref{fig:FIG00004}(e), and~\ref{fig:FIG00004}(h)]. In particular, for $\alpha = 0$ and $\alpha = 1$, with $\theta = \{\pi/4,\pi/2\}$, we have that the entropic QSL approaches small values at earlier ($\Gamma\tau \ll 1$) and later ($\Gamma\tau \gg 1$) times of the dynamics for initial mixed states, regardless $0 < r < 1$ [see Figs.~\ref{fig:FIG00004}(d),~\ref{fig:FIG00004}(f),~\ref{fig:FIG00004}(g), and~\ref{fig:FIG00004}(i)]. In this setting, one finds that ${\tau_J^{\text{QSL}}}$ exhibits a nonmonotonic behavior as a function of $\Gamma\tau$ along the nonunitary evolution of the two-level system. Indeed, Figs.~\ref{fig:FIG00004}(d),~\ref{fig:FIG00004}(f),~\ref{fig:FIG00004}(g), and~\ref{fig:FIG00004}(i) show that, for fixed value $r \in [0,1)$, the QSL time starts growing but smoothly approaches small values as a function of $\Gamma\tau \geq 0$. It is noteworthy that, for $\alpha = \{0,1\}$ and $\theta = \pi/2$, ${\tau_J^{\text{QSL}}}$ [see Eq.~\eqref{eq:0000026}] exhibits the same quantitative behavior, which can be numerically verified in Figs.~\ref{fig:FIG00004}(g) and~\ref{fig:FIG00004}(i).

In Fig.~\ref{fig:FIG00005}, we show the density plots of the normalized relative error ${\widetilde{\delta}_{J}}({\tau})$, as a function of the dimensionless parameters $\Gamma\tau$ and $r$. The entropic QSL bound is loose at earlier times of the dynamics, regardless of the initial state and the value of the parameter $\alpha \in [0,1]$. In addition, Figs.~\ref{fig:FIG00005}(c) and~\ref{fig:FIG00005}(f) show that the bound is also loose for probe states close to the incoherent state $|0\rangle\langle{0}|$, for all $\Gamma\tau \geq 0$. Overall, for $\alpha \ll 1$ and $\theta = \{0,\pi/4,\pi/2\}$, one finds that ${\widetilde{\delta}_{J}}({\tau})$ approaches small values for all $\Gamma\tau \gg 1$ [see Figs.~\ref{fig:FIG00005}(a),~\ref{fig:FIG00005}(b),~\ref{fig:FIG00005}(d),~\ref{fig:FIG00005}(e),~\ref{fig:FIG00005}(g), and~\ref{fig:FIG00005}(h)]. In particular, for $\theta = \pi/2$, tighter entropic QSLs are found at later times of the dynamics whenever the initial state is close to the maximally mixed state, regardless the value of $\alpha = \{0,0.1,1\}$ [see Figs.~\ref{fig:FIG00005}(g),~\ref{fig:FIG00005}(h), and~\ref{fig:FIG00005}(i)].

In Fig.~\ref{fig:FIG00006}, we show the density plots of the entropic QSL time ${\tau_{JS}^{\text{QSL}}}$, as a function of the dimensionless parameters $\Gamma\tau$ and $r$. We note that, for a fixed value $r \in [0,1)$ of the mixedness parameter, ${\tau_{JS}^{\text{QSL}}}$ varies monotonically as a function of the dimensionless parameter $\Gamma\tau \geq 0$. Overall, the entropic QSL approaches small values at earlier times of the dynamics, exhibiting similar qualitative behavior regardless the initial state and temperature of the thermal bath that we consider here. In particular, for $\alpha = \{0,1\}$ and $\theta = \pi/2$, Figs.~\ref{fig:FIG00006}(g) and~\ref{fig:FIG00006}(i) show that the QSL time exhibits the same quantitative behavior. This comes from the fact that ${\tau_{JS}^{\text{QSL}}}$ reduces to the same result under this choice of parameters, regardless of $r \in [0,1)$ and $\Gamma\tau \geq 0$ [see Eq.~\eqref{eq:0000042}]. This means that, for a given probe mixed state lying in the equatorial plane of the Bloch sphere, with $0 < r < 1$ and $\theta = \pi/2$, the QSL time at which the system reaches the stationary states $|1\rangle\langle{1}|$ ($\alpha = 0$) or $|0\rangle\langle{0}|$ ($\alpha = 1$) will be the same.

In Fig.~\ref{fig:FIG00007}, we show the density plots of the normalized relative error ${\widetilde{\delta}_{JS}}({\tau})$, as a function of the dimensionless parameters $\Gamma\tau$ and $r$. We note that, for a fixed value $r \in [0,1)$ of the mixedness parameter, ${\widetilde{\delta}_{JS}}({\tau})$ monotonically decreases as a function of the dimensionless parameter $\Gamma\tau \geq 0$. In particular, for $\Gamma\tau \lesssim 0.25$, one finds that $0.8 \lesssim {\widetilde{\delta}_{JS}}({\tau}) \lesssim 1$, with $r \in [0,1)$. This means that the entropic QSL bound is looser at earlier times of the dynamics, regardless of the purity of the initial single-qubit state. Furthermore, we have that ${\widetilde{\delta}_{JS}}({\tau})$ approaches small values at later times of the dynamics. For example, Figs.~\ref{fig:FIG00007}(c),~\ref{fig:FIG00007}(f),~\ref{fig:FIG00007}(g),~\ref{fig:FIG00007}(h), and~\ref{fig:FIG00007}(i) show that ${\widetilde{\delta}_{JS}}({\tau}) \lesssim 0.2$ for $0 \leq r \lesssim 0.2$ and $\Gamma\tau \gtrsim 2$, with $\theta = \{0,\pi/4,\pi/2\}$ and $\alpha = \{0,0.1,1\}$.


\section{Tightness of QSL bounds}
\label{sec:00000000006}

\begin{figure}[!t]
\begin{center}
\includegraphics[scale=0.9]{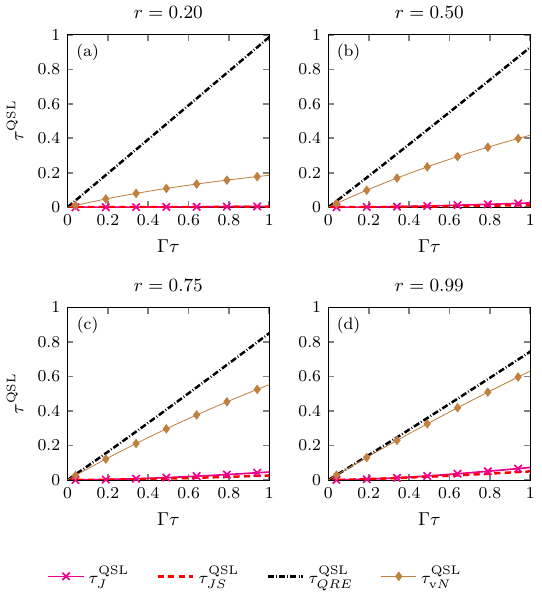}
\caption{(Color online) Plot of the QSL time $\tau^{\text{QSL}}$ as a function of the dimensionless parameter $\Gamma\tau$, for a single-qubit state evolving under the depolarizing channel. We consider the QSLs ${\tau^{\text{QSL}}_J}$ (magenta solid line with cross markers), ${\tau^{\text{QSL}}_{JS}}$ (red dashed line), ${\tau^{\text{QSL}}_{QRE}}$ (black dash-dotted line), and ${\tau^{\text{QSL}}_{\text{v}N}}$ (brown solid line with diamond markers). Here we set initial single-qubit states with mixedness parameters (a) $r = 0.20$, (b) $r = 0.50$, (c) $r = 0.75$, and (d) $r = 0.99$, for all $\theta \in [0,\pi]$ and $\phi \in [0,2\pi)$.}
\label{fig:FIG00008}
\end{center}
\end{figure}

In this section, we investigate the tightness of our QSL bounds. To so, we compare their performance to other proposed entropic quantum speed limits. We focus on the QSLs in Eqs.~\eqref{eq:0000016} and~\eqref{eq:0000017} for nonunitary dynamics, in comparison with the entropic speed limits derived in Refs.~\cite{NewJPhys_24_065003,PhysRevA.107.052419}. On the one hand, Ref.~\cite{PhysRevA.107.052419} presented a QSL based on Umegaki relative entropy for open quantum systems, based on the Schatten 2-norm of a time-dependent Liouvillian operator ${\mathcal{L}_t}(\bullet)$ governing the nonunitary dynamics, given by $\tau \geq {\tau_{QRE}^{\text{QSL}}}$, with
\begin{equation}
\label{eq:0000050}
{\tau_{QRE}^{\text{QSL}}} := \frac{S({\rho_{\tau}}\|{\rho_0})}{\langle\langle{\|{\mathcal{L}_t}({\rho_t})\|_2}{\|\ln{\rho_t} - \ln{\rho_0}\|_2}\rangle\rangle_{\tau}} ~,
\end{equation}
where ${\langle\langle\bullet\rangle\rangle_{\tau}} := (1/\tau){\int_0^{\tau}} dt \,\bullet$ is the time average, ${\|X\|_2} = \sqrt{\text{Tr}({X^{\dagger}}X)}$, and $d{\rho_t}/dt = {\mathcal{L}_t}({\rho_t})$, for all $t \in [0,\tau]$. On the other hand, Ref.~\cite{NewJPhys_24_065003} introduced a QSL time based on the von Neumann entropy for open quantum systems, which also depends on the Schatten 2-norm of the Liouvillian operator of the dynamics, which is given by $\tau \geq {\tau_{\text{v}N}^{\text{QSL}}}$, with
\begin{equation}
\label{eq:0000051}
{\tau_{\text{v}N}^{\text{QSL}}} := \frac{|S({\rho_{\tau}}) - S({\rho_0})|}{\sqrt{{\langle\langle{\|{\mathcal{L}_t}({\rho_t})\|_2^2}\rangle\rangle_{\tau}}{\langle\langle{\|\ln{\rho_t}\|_2^2}\rangle\rangle_{\tau}}}} ~.
\end{equation}
The QSLs in Eqs.~\eqref{eq:0000050} and~\eqref{eq:0000051} depend on Schatten 2-norm of the generator of the nonunitary dynamics, whose evaluation requires the complete spectral decomposition of the Liouvillian of the open quantum system, as well as the eigenvalues of the initial and instantaneous states. In contrast, Eqs.~\eqref{eq:0000016} and~\eqref{eq:0000017} are simpler to compute while still providing meaningful lower bounds on the evolution time of many-particle systems. These QSLs depend on the Kraus operators governing the nonunitary evolution, as well as the smallest and largest eigenvalues of the initial and instantaneous states. Notably, for closed quantum systems, Eq.~\eqref{eq:0000051} yields a trivial bound since the von Neumann entropy is conserved under unitary evolutions, whereas our results apply equally to both closed and open systems. Furthermore, the entropic speed limits in Eqs.~\eqref{eq:0000016} and~\eqref{eq:0000017} depend on the distance traveled in the space of quantum states, as captured by both the QJPD and QJSD, thereby providing a geometric perspective on the quantum state evolution. These bounds also convey information about the quantum coherences of the evolved state relative to the eigenstates of the probe state.

\begin{figure}[!t]
\begin{center}
\includegraphics[scale=0.9]{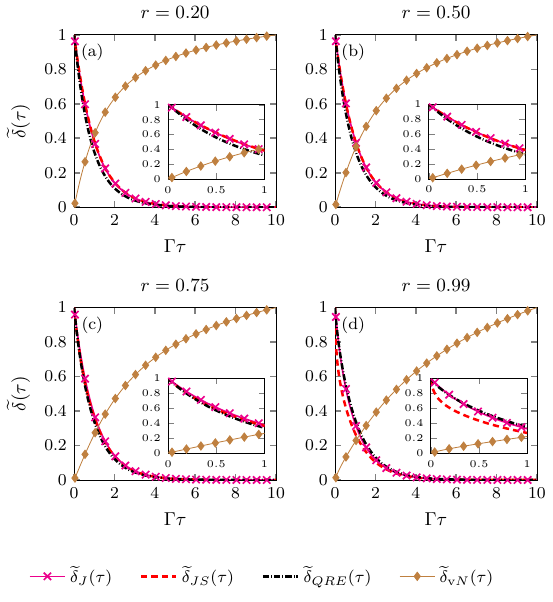}
\caption{(Color online) Plot of the normalized relative error ${\widetilde{\delta}}(\tau)$ as a function of the dimensionless parameter $\Gamma\tau$, for a single-qubit state evolving under the depolarizing channel, with $0 \leq \Gamma\tau \leq 10$. We consider the relative errors ${\widetilde{\delta}_J}(\tau)$ (magenta solid line with cross markers), ${\widetilde{\delta}_{JS}}(\tau)$ (red dashed line), ${\widetilde{\delta}_{QRE}}(\tau)$ (black dash-dotted line), and ${\widetilde{\delta}_{\text{v}N}}(\tau)$ (brown solid line with diamond markers). Here we set initial single-qubit states with mixedness parameters (a) $r = 0.20$, (b) $r = 0.50$, (c) $r = 0.75$, and (d) $r = 0.99$, for all $\theta \in [0,\pi]$ and $\phi \in [0,2\pi)$. The insets show the normalized relative error at earlier times of the dynamics.}
\label{fig:FIG00009}
\end{center}
\end{figure}

In the following, we specialize Eqs.~\eqref{eq:0000016},~\eqref{eq:0000017},~\eqref{eq:0000050}, and~\eqref{eq:0000051} to the case of a single-qubit state evolving nonunitarily under the depolarizing channel, thus providing numerical simulation of the bounds. The details of the dynamics are given in Sec.~\ref{sec:00000000005B1}. In Fig.~\ref{fig:FIG00008}, we plot the QSL bounds ${\tau^{\text{QSL}}_J}$ (magenta solid line cross markers), ${\tau^{\text{QSL}}_{JS}}$ (red dashed line), ${\tau^{\text{QSL}}_{QRE}}$ (black dash-dotted line), and ${\tau^{\text{QSL}}_{\text{v}N}}$ (brown solid line with diamond markers), as a function of the dimensionless parameter $\Gamma\tau$, with $0 \leq \Gamma\tau \leq 1$. For completeness, Fig.~\ref{fig:FIG00009} shows the normalized relative errors ${\widetilde{\delta}_J}(\tau)$ (magenta solid line cross markers), ${\widetilde{\delta}_{JS}}(\tau)$ (red dashed line), ${\widetilde{\delta}_{QRE}}(\tau)$ (black dash-dotted line), and ${\widetilde{\delta}_{\text{v}N}}(\tau)$ (brown solid line with diamond markers), as a function of the dimensionless parameter $\Gamma\tau$, with $0 \leq \Gamma\tau \leq 10$. In each panel, the inset shows the normalized relative error at earlier times of the dynamics, as a function of $\Gamma\tau$, with $0 \leq \Gamma\tau \leq 1$. In both figures, we set initial single-qubit states respective to the mixedness parameters: $r = 0.20$ [Figs.~\ref{fig:FIG00008}(a) and~\ref{fig:FIG00009}(a)], $r = 0.50$ [Figs.~\ref{fig:FIG00008}(b) and~\ref{fig:FIG00009}(b)], $r = 0.75$ [Figs.~\ref{fig:FIG00008}(c) and~\ref{fig:FIG00009}(c)], and $r = 0.99$ [Figs.~\ref{fig:FIG00008}(d) and~\ref{fig:FIG00009}(d)].

From Figs.~\ref{fig:FIG00008} and~\ref{fig:FIG00009}, the bounds in Eqs.~\eqref{eq:0000016} and~\eqref{eq:0000017} [see also Eqs.~\eqref{eq:0000030} and~\eqref{eq:0000036}] are looser compared with Eqs.~\eqref{eq:0000050} and~\eqref{eq:0000051} for the considered dynamics. The inset in Fig.~\ref{fig:FIG00009}(d) shows that, at early times, the performance of ${\tau^{\text{QSL}}_{J,JS}}$ slightly improves for initial states closer to the surface of the Bloch sphere, yet remains looser than ${\tau^{\text{QSL}}_{QRE}}$ at later times of the dynamics. Next, we discuss the operational and physical features related to this discrepancy. For a single-qubit evolving under the action of the depolarizing quantum channel, the bounds given in Eqs.~\eqref{eq:0000016} and~\eqref{eq:0000017} attain significantly smaller magnitudes than those in Eqs.~\eqref{eq:0000050} and~\eqref{eq:0000051}. Specifically, the term ${\sum_j}\,{\|{K_j}{\rho_0}(d{K^{\dagger}_j}/dt)\|_1}$ decreases monotonically with time, independently of the mixedness $r$ of the initial state [see Table~\ref{tab:TABLE01}]. The cost functions ${f_{J,JS}}({\rho_0},{\rho_t})$ decrease with time for fixed $r$, while they increase with $r$ for fixed time. Consequently, the time-averaged quantities ${\langle\langle{f_{J,JS}}({\rho_0},{\rho_t}){\sum_j}\,{\|{K_j}{\rho_0}(d{K^{\dagger}_j}/dt)\|_1}\rangle\rangle_{\tau}}$ decrease monotonically with time for fixed mixedness, approaching small values at longer times of the dynamics. In turn, the QJPD [see Eq.~\eqref{eq:0000027}] and QJSD [see Eq.~\eqref{eq:0000033}] increase monotonically with time for fixed mixedness and saturate at longer times of the dynamics [see Eqs.~\eqref{eq:0000029} and~\eqref{eq:0000035}], also exhibiting small magnitudes. Both entropic measures depend on the mixedness of the initial state but are independent of the angular parameters $(\theta,\phi)$, which is consistent with the isotropic action of the depolarizing channel that uniformly contracts the Bloch sphere. As a result, the corresponding entropic QSLs become increasing functions of time, exhibiting linear growth at longer times, while remaining smaller than the bounds in Eqs.~\eqref{eq:0000050} and~\eqref{eq:0000051}. We obtain QSLs of order ${\tau^{\text{QSL}}_{J,JS}} \sim {10^{-3}}$ for $r = 0.20$ [see Fig.~\ref{fig:FIG00008}(a)], and ${\tau^{\text{QSL}}_{J,JS}} \sim {10^{-2}}$ for $r = \{0.50, 0.75, 0.99\}$ [see Figs.~\ref{fig:FIG00008}(b), \ref{fig:FIG00008}(c), and \ref{fig:FIG00008}(d), respectively], within the interval $\Gamma\tau \in [0,1]$. This accounts for the smaller magnitudes of the bounds in Eqs.~\eqref{eq:0000016} and~\eqref{eq:0000017} and is consistent with the behavior observed in our simulations.

The simulations further indicate that the number of inequalities used in deriving the entropic bounds may compromise their tightness. As illustrated in Figs.~\ref{fig:FIG00008} and~\ref{fig:FIG00009}, the entropic QSLs and their corresponding normalized relative errors display poorer performance than the bounds in Eqs.~\eqref{eq:0000050} and~\eqref{eq:0000051} for the depolarizing nonunitary dynamics of single-qubit states. Tighter bounds could be obtained by reducing the number of inequalities, though the evaluation of such bounds may become considerably more involved, particularly for higher-dimensional systems. We also note that the bounds in Eqs.~\eqref{eq:0000050} and~\eqref{eq:0000051} are derived through a distinct methodology, based on a different combination of matrix-norm inequalities, which naturally leads to a different family of QSLs whose tightness is intrinsically linked to this construction. Finally, since the QSL time depends sensitively on (i) the initial state of the physical system, and (ii) the dynamical process, general statements about the relative advantage of the bounds require systematic analysis over a broader class of quantum channels. We hope that these results will stimulate further investigations in this direction.


\section{Conclusions}
\label{sec:00000000007}

In this work, we have addressed quantum speed limits based on quantum versions of the Jeffreys divergence and Jensen-Shannon divergence. These distinguishability measures correspond to symmetrizations of the standard quantum relative entropy, also called Umegaki relative entropy. Both are nonnegative, symmetric with respective to the swap of the two input states, contractive under the action of completely positive and trace-preserving maps, but do not satisfy triangle inequality. So far, the square root of the Jensen-Shannon divergence, called the quantum Jensen-Shannon distance (QJSD), has proven to be a \textit{bona fide} information metric on the convex manifold of quantum states, but there is no clear evidence that the square root of the Jeffreys divergence, called the quantum Jeffreys pseudodistance (QJPD), also shares this property. We study QJPD and QJSD to obtain entropic QSLs that apply to finite-dimensional quantum systems undergoing general physical processes.

We investigate the dynamical behavior of QJPD and QJSD evaluated for states of arbitrary quantum systems, thus proving upper bounds on these quantities [see Eq.~\eqref{eq:0000007}]. The bounds are evaluated in terms of the Schatten speed of the instantaneous state, as well as cost functions that depend on the smallest and largest eigenvalues of both initial and instantaneous states of the quantum system [see Eqs.~\eqref{eq:0000008} and~\eqref{eq:0000009}]. These bounds are expected to require low computational efforts, mainly because their evaluation involves few eigenvalues of the probe and evolved states, thus providing useful estimates of QJSD and QJPD for higher-dimensional quantum systems.

We address entropic QSLs based on the QJPD [see Eq.~\eqref{eq:0000010}] and QJSD [see Eq.~\eqref{eq:0000013}], which apply for both closed and open quantum systems. On the one hand, for unitary dynamics, we obtain QSLs from QJPD [see Eq.~\eqref{eq:0000014}] and QJSD [see Eq.~\eqref{eq:0000015}] that depend on the Schatten $1$-norm of the commutator between the probe state and the Hamiltonian that generates the dynamics. This quantity signals the role played by the coherences of the initial state with respect to the fixed eigenbasis of the Hamiltonian. In this setting, one readily verifies that the QSLs are inversely proportional to the variance of the time-independent Hamiltonian, i.e., our results fit into the class of Mandelstam-Tamm speed limits. On the other hand, for nonunitary evolutions described by CPTP maps, we derive entropic QSLs from QJPD [see Eq.~\eqref{eq:0000016}] and QJSD [see Eq.~\eqref{eq:0000017}] written in terms of the Kraus operators that describes the overall evolution.

To illustrate our findings, we have specialized these results to the case of two-level systems undergoing (i) unitary dynamics [see Sec.~\ref{sec:00000000005A}], and (ii) nonunitary dynamics [see Sec.~\ref{sec:00000000005B}]. On the one hand, for closed quantum systems, we consider the unitary evolution of mixed single-qubit states generated by a time-independent Hamiltonian. We obtain analytical QSLs based on the QJPD [see Eq.~\eqref{eq:0000020}] and QJSD [see Eq.~\eqref{eq:0000022}], also discussing their dynamical properties. On the other hand, for open quantum systems, we set the nonunitary dynamics generated by three quantum operations, namely, depolarizing channel, phase damping channel, and generalized amplitude damping channel. We provide analytical results for the QJPD and QJSD in each case, and also present numerical simulations to support our fin\-dings about QSLs [see Figs.~\ref{fig:FIG00002},~\ref{fig:FIG00003},~\ref{fig:FIG00004},~\ref{fig:FIG00006}, and~\ref{fig:FIG00008}]. We also address the tightness of these speed limits [see Figs.~\ref{fig:FIG00002},~\ref{fig:FIG00003},~\ref{fig:FIG00005},~\ref{fig:FIG00007}, and~\ref{fig:FIG00009}]. The bounds are looser at earlier times of the dynamics, and they are not tight for the overall evolution.

It is worth mentioning that entropic QSLs based on $\chi$-$z$-R\'{e}nyi relative entropies were recently presented in Ref.~\cite{ghr4-d2vb}. Hence, it is natural to ask in what extent such results can or can not overcome the bounds discussed here, particularly since it would be possible to recover Umegaki's relative entropy from $\chi$-$z$-R\'{e}nyi relative entropies by choosing $z = 1$, and taking the limit $\chi \rightarrow 1$. We point out that the results discussed here are not compatible with those addressed in Ref.~\cite{ghr4-d2vb}. First, the results in Ref.~\cite{ghr4-d2vb} does apply only for $\chi \in (0,1)$ and $1 \geq z \geq \text{max}\{\chi, 1 - \chi\}$, with $\chi \neq 1$, which rules out any chance of obtaining QSLs based on standard relative entropy from $\chi$-$z$-R\'{e}nyi relative entropies. This means that the bounds presented here cannot be recovered from the results in Ref.~\cite{ghr4-d2vb}, neither as particular cases nor as straightforward applications. Second, the number of inequalities used can actually lead to different families of bounds, which also undermines the tightness of these quantities. Indeed, tighter bounds on the QJPD and QJSD are expected by applying a small number of inequalities, which may imply tighter and attainable speed limits. Leveraging the sensitivity of our QSLs respective to a few eigenvalues of the system overcomes the need to evaluate the full spectrum of effective Liouvillians in open quantum systems, which is the case of Refs.~\cite{NewJPhys_24_065003,arXiv:2303.07415,PhysRevA.107.052419} that showed QSLs based on relative entanglement entropies.

We emphasize that our bounds require minimal information about the physical system, e.g., the smallest and largest eigenvalues of the initial and evolved states, and also details about the generator of the dynamics. In the following, we comment on three experimental proposals for investigating our speed limits. First, the entropic QSLs for the nonunitary dynamics of single-qubit states can be experimentally validated using nuclear spins in nuclear magnetic resonance (NMR) setups, according to Ref.~\cite{arXiv:2307.06558}. In this setup, using chloroform molecules ($\text{CHCl}_3$), the qubits are synthesised in the carbon nuclear spins ($^{13}$C) and experience decoherence from the hydrogen nuclear spins ($^{1}$H). External radio-frequency pulses control the spin precession, enabling the observation of longitudinal and transverse magnetization dynamics during relaxation. Notably, both the QJPD and QJSD can be recast in terms of these magnetizations, since they are directly linked to the coherences and populations of the instantaneous single-qubit state. Second, the entropic bounds could be experimentally analyzed using moments of the energy spectra and quantum state tomography of the unitary dynamics of transmon qubits in superconducting circuits, as discussed Ref.~\cite{PhysRevA.110.042215}. In this setting, Ref.~\cite{NatCommun_16_1255_2025} provides a complementary experimental setup to probe QSLs for the unitary dynamics of qubits, qutrits, and multi-qubit systems under a tunable external microwave drive. The bounds could be experimentally accessed by expressing the QJPD and QJSD in terms of the overlap between initial and instantaneous states of the dynamics, which are determined from the probability oscillations measured in the experiment. Finally, we note that entropic QSLs could also be experimentally tested with the interferometric setup described in Ref.~\cite{Photonics_10_1004_2023}, where qubit states are encoded in the polarization degrees of freedom of photons. The approach involves relating the probability distribution of measurement outcomes to the maximum interference visibility of a Mach-Zehnder interferometer, using quantum state tomography. To compare experimental data with numerical simulations, we recast both the QJPD and QJSD using the spectral decomposition of the qubit states, expressing them in terms of the experimentally obtained probability outcomes. Therefore, we rewrite the entropic bounds in terms of such probabilities, which allow us compare the experimental results with the respective numerical simulation of our speed limits.

Our results can be useful in the study of entropic uncertainty relations~\cite{arXiv:2203.12421,QSL_thermod001}, particularly based on Jensen-Shannon and Jeffreys divergences. These QSLs may also find applications in quantum thermodynamics and work fluctuations~\cite{PhysRevLett.123.230603,PhysRevResearch.2.023377,5lp2-9sps}, noisy quantum metrology~\cite{QSL_metrology001}, study of characterization of non-Markovianity~\cite{PhysRevLett.127.030401,PhysRevA.106.042212}, study of magic resources of quantum states~\cite{JPhysA_58_015303_2024}, quantum phase transitions~\cite{PhysicaA_561_125176_2021}, and also in the study of complexity in quantum many-body quantum dynamics~\cite{ycdh-z8zf}.


\begin{acknowledgments}
This work was supported by the Brazilian ministries MEC and MCTIC, and the Brazilian funding agencies CNPq, and Coordena\c{c}\~{a}o de Aperfei\c{c}oamento de Pessoal de N\'{i}vel Superior--Brasil (CAPES) (Finance Code 001). D. P. P. would like to acknowledge the Funda\c{c}\~{a}o de Amparo \`{a} Pesquisa e ao Desenvolvimento Cient\'{i}fico e Tecnol\'{o}gico do Maranh\~{a}o (FAPEMA) (Edital Acordo de Coopera\c{c}\~{a}o T\'{e}cnica - Bolsas Produtividade em Pesquisa Estaduais FAPEMA/CNPq, Grant No.~PQ-C-12651/25).
\end{acknowledgments}

\setcounter{equation}{0}
\setcounter{table}{0}
\setcounter{section}{0}
\numberwithin{equation}{section}
\makeatletter
\renewcommand{\thesection}{\Alph{section}} 
\renewcommand{\thesubsection}{\Alph{section}.\arabic{subsection}}
\def\@gobbleappendixname#1\csname thesubsection\endcsname{\Alph{section}.\arabic{subsection}}
\renewcommand{\theequation}{\Alph{section}\arabic{equation}}
\renewcommand{\thefigure}{\arabic{figure}}
\renewcommand{\bibnumfmt}[1]{[#1]}
\renewcommand{\citenumfont}[1]{#1}

\section*{Appendix}


\section{Proof of Eq.~$\eqref{eq:0000007}$}
\label{sec:0000000000A000}

In this Appendix, we provide the detailed proof of Eq.~\eqref{eq:0000007} for the Jeffreys pseudodistance, and Jensen-Shannon distance.


\subsection{Bounding quantum Jeffreys pseudodistance}
\label{sec:00000000003A}

We consider the QJPD, ${D_J}({\rho_0},{\rho_t})$, as a symmetric distinguishability measure of the quantum states ${\rho_0}$ and ${\rho_t}$ [see Eq.~\eqref{eq:0000003}]. The absolute value of the time derivative of the squared QJPD satisfies the upper bound
\begin{align}
\label{eq:appendixA0001}
& \left|\frac{d}{dt}{D_J^2}({\rho_0},{\rho_t})\right| \leq \frac{1}{2}\left|\frac{d}{dt}S({\rho_t})\right| \nonumber\\
& + \frac{1}{2}\left|\text{Tr}\left({\rho_0}\frac{d}{dt}\ln{\rho_t}\right)\right| + \frac{1}{2}\left|\text{Tr}\left(\ln{\rho_0}\frac{d{\rho_t}}{dt}\right)\right| ~,
\end{align}
where we have applied the triangle inequality, i.e., $|{A_1} + {A_2}| \leq |{A_1}| + |{A_2}|$. In Appendix~\ref{sec:0000000000C}, we prove that the time derivative of von Neumann entropy becomes
\begin{equation}
\label{eq:appendixA0002}
\frac{d}{dt}S({\rho_t}) = -\text{Tr}\left(\ln{\rho_t}\frac{d{\rho_t}}{dt}\right) ~,
\end{equation}
which holds for an arbitrary time-dependent density matrix ${\rho_t} = {\mathcal{E}_t}(\rho_0) \in \mathcal{S}$ that undergoes a given physical process. In addition, in Appendix~\ref{sec:0000000000D} we prove that
\begin{equation}
\label{eq:appendixA0003}
\left| \frac{d}{dt}\text{Tr}({\rho_0}\ln{\rho_t}) \right| \leq \frac{{\kappa_{\text{max}}}({\rho_0})}{{\kappa_{\text{min}}}({\rho_t})} \, {\left\|\frac{d{\rho_t}}{dt}\right\|_1} ~.
\end{equation}
By combining Eqs.~\eqref{eq:appendixA0001},~\eqref{eq:appendixA0002}, and~\eqref{eq:appendixA0003}, we conclude that the rate of change of the squared QJPD satisfies the upper bound
\begin{equation}
\label{eq:appendixA0004}
\left|\frac{d}{dt}{D_J^2}({\rho_0},{\rho_t})\right| \leq {f_J}({\rho_0},{\rho_t}){\left\|\frac{d{\rho_t}}{dt}\right\|_1} ~,
\end{equation}
where the auxiliary function ${f_J}({\rho_0},{\rho_t})$ is defined in Eq.~\eqref{eq:0000008}. We also use the fact that $|\text{Tr}({A_1}{A_2})| \leq {\|{A_1}\|_{\infty}}{\|{A_2}\|_1}$, with ${\|\ln\bullet\|_{\infty}} = |\ln\left({\kappa_{\text{min}}}(\bullet)\right)|$. To obtain an upper bound on QJPD, we integrate Eq.~\eqref{eq:appendixA0004} over the time interval $t\in[0,\tau]$, thus obtaining the upper bound
\begin{equation}
\label{eq:appendixA0005}
{D_J}({\rho_0},{\rho_{\tau}}) \leq \sqrt{{\int_0^{\tau}} {dt} \, {f_J}({\rho_0},{\rho_t}) \, {\left\|\frac{d{\rho_t}}{dt}\right\|_1}} ~,
\end{equation}
where we have applied the inequality $\left|\int d{t} \, g(t)\right| \leq \int d{t} \, |g(t)|$, and we also extract the square root of both sides of the resulting inequality.

Based on Pinsker's inequality, it is known that quantum Jeffreys divergence satisfies the lower bound ${S_J}({\rho_0}\|{\rho_{\tau}}) \geq (1/2){\|{\rho_0} - {\rho_{\tau}}\|_1^2}$. In this case, from Eq.~\eqref{eq:appendixA0005}, one finds the chain of inequalities on QJPD as
\begin{equation}
\label{eq:appendixA0006}
\frac{{\|{\rho_0} - {\rho_{\tau}}\|_1}}{\sqrt{2}} \leq {D_J}({\rho_0},{\rho_{\tau}}) \leq \sqrt{{\int_0^{\tau}} {dt} \, {f_J}({\rho_0},{\rho_t}) \, {\left\|\frac{d{\rho_t}}{dt}\right\|_1}} ~.
\end{equation}
Overall, Eq.~\eqref{eq:appendixA0006} provides a hierarchical bound on QJPD. On the one hand, QJPD is bounded from below by the trace distance between initial and final states. On the other hand, QJPD is upper bounded by the time-averaged Schatten speed along the evolution path followed by the instantaneous quantum state.


\subsection{Bounding quantum Jensen-Shannon distance}
\label{sec:00000000003B}

Next, we present an upper bound on the square root of the quantum Jensen-Shannon divergence, called the quantum Jensen-Shannon distance (QJSD), which in turn defines a metric on the space of quantum states. The absolute value of the time derivative of QJSD [see Eq.~\eqref{eq:0000006}] goes as follows
\begin{align}
\label{eq:appendixA0007}
\left|\frac{d}{dt}{D_{JS}^2}({\rho_0},{\rho_t})\right| \leq \left|\frac{d}{dt}S({\varpi_t})\right| + \frac{1}{2}\left|\frac{d}{dt}S({\rho_t})\right| ~,
\end{align}
where we have applied the triangular inequality, $|{A} + {B}| \leq |{A}| + |{B}|$, with ${\varpi_t} := (1/2)({\rho_0} + {\rho_t})$. We note that ${\varpi_t} \in \mathcal{S}$, with ${\varpi_t^{\dagger}} = {\varpi_t}$, $\text{Tr}({\varpi_t}) = 1$, and ${\varpi_t} \geq 0$. From Eq.~\eqref{eq:appendixA0002}, one finds the rate of change $dS({\varrho_t})/dt = -\text{Tr}\left[(d{\varrho_t}/dt)\ln{\varrho_t}\right]$ of the von Neumann entropy for an arbitrary time-dependent density matrix $\varrho_t \in \mathcal{S}$ [see also Appendix~\ref{sec:0000000000C}]. In this case, by using the matrix norm inequality $|\text{Tr}({A_1}{A_2})| \leq {\|{A_1}\|_{\infty}}{\|{A_2}\|_1}$, we conclude that the rate of change of the von Neumann entropy satisfies the upper bound
\begin{equation}
\label{eq:appendixA0008}
\left|\frac{d}{dt}S({\varrho_t})\right| \leq {\|\ln{\varrho_t}\|_{\infty}}{\left\|\frac{d{\varrho_t}}{dt}\right\|_1} ~.
\end{equation}
By combining Eqs.~\eqref{eq:appendixA0007} and~\eqref{eq:appendixA0008}, we arrive at the result
\begin{equation}
\label{eq:appendixA0009}
\left|\frac{d}{dt}{D_{JS}^2}({\rho_0},{\rho_t})\right| \leq \frac{1}{2}\left( {\|\ln{\varpi_t}\|_{\infty}} + {\|\ln{\rho_t}\|_{\infty}}\right){\left\|\frac{d{\rho_t}}{dt}\right\|_1} ~.
\end{equation}
Next, by integrating Eq.~\eqref{eq:appendixA0009} over the time interval $t\in [0,\tau]$, one gets
\begin{equation}
\label{eq:appendixA0010}
{D_{JS}}\left(\rho_{0},\rho_{\tau}\right) \leq \sqrt{{\int_0^{\tau}}dt\, {f_{JS}}({\rho_0},{\rho_t}) \, {\left\|\frac{d{\rho_t}}{dt}\right\|_1}} ~,
\end{equation}
where we have used the inequality $\left|\int d{t} \, g(t)\right| \leq \int d{t} \, |g(t)|$, with $|\ln(xy)| = |\ln{x}| + |\ln{y}|$ for all $0 < x \leq 1$ and $0 < y \leq 1$, and we reinforce that the auxiliary function is defined in Eq.~\eqref{eq:0000009}. This concludes our proof.


\section{Logarithm of density matrices}
\label{sec:0000000000B}

In this Appendix, we address the proof of the integral representation of the logarithm of density matrices as follows 
\begin{equation}
\label{eq:appendixB0001}
\ln{\rho} = {\int_0^{\infty}} du \left(\frac{1}{1 + u}\mathbb{I} - {(\rho + u\mathbb{I})^{-1}}\right) ~,
\end{equation}
where $\mathbb{I}$ is the identity matrix, and $\mathcal{S} = \{\rho\in\mathcal{H} \mid {\rho^{\dagger}} = \rho,~\rho \geq 0,~\text{Tr}(\rho) = 1\}$ defines the convex subspace of a $d$-dimensional Hilbert space $\mathcal{H}$, with $d = \dim\mathcal{H}$. To do so, let $\rho = {\sum_j}\,{p_j}|j\rangle\langle{j}|$ be a nonsingular, full rank density matrix, where $\{{p_j}\}_{j = 1,\ldots,d}$ and $\{|j\rangle\}_{j = 1,\ldots,d}$ define the sets of eigenvalues and eigenstates, respectively. We note that $0 < {p_j} < 1$ and ${\sum_j}\,{p_j} = 1$, $\langle{j}|{\ell}\rangle = {\delta_{j\ell}}$ and ${\sum_j}|j\rangle\langle{j}| = \mathbb{I}$, where $\mathbb{I}$ is the identity matrix. Hence, by exploiting the aforementioned spectral decomposition of the density matrix, one gets
\begin{equation}
\label{eq:appendixB0002}
\ln\rho = {\sum_j} \,\ln{p_j}|j\rangle\langle{j}| ~.
\end{equation}
We note that the function $\ln{p_j}$ exhibits a monotonic behavior as a function of the parameter $0 < {p_j} < 1$, for all $j = \{1,\ldots,d\}$, and can be written as in terms of the following representation~\cite{Bathia,ProgTheoPhys_100_475_1998}
\begin{equation}
\label{eq:appendixB0003}
\ln{p_j} = {\int_0^{\infty}} du \left(\frac{1}{1 + u} - \frac{1}{{p_j} + u}\right) ~.
\end{equation}
Hence, by combining Eqs.~\eqref{eq:appendixB0002} and~\eqref{eq:appendixB0003}, it yields
\begin{align}
\label{eq:appendixB0004}
\ln\rho &= {\int_0^{\infty}} du \, \left(\frac{1}{1 + u} \, {\sum_j}\, |j\rangle\langle{j}| - {\sum_j} \frac{1}{{p_j} + u} |j\rangle\langle{j}| \right) \nonumber\\
&= {\int_0^{\infty}} du \, \left(\frac{1}{1 + u}\mathbb{I} - {\sum_j} \, \frac{1}{{p_j} + u}|j\rangle\langle{j}| \right) ~,
\end{align}
where we have used the completeness relation of the set of eigenstates of the density matrix. It can also be proved that 
\begin{align}
\label{eq:appendixB0005}
{(\rho + u\mathbb{I})^{-1}} &= {\left({\sum_j}\, ({p_j} + u)|j\rangle\langle{j}|\right)^{-1}} \nonumber\\
&=  {\sum_j} \, \frac{1}{{p_j} + u}|{j}\rangle\langle{j}| ~.
\end{align}
Finally, combining Eqs.~\eqref{eq:appendixB0004} and~\eqref{eq:appendixB0005}, also re\-cogni\-zing the spectral decomposition of the density matrix, we conclude the integral representation of the logarithm of density matrices given in Eq.~\eqref{eq:appendixB0001}.


\section{Proof of Eq.~$\eqref{eq:appendixA0002}$}
\label{sec:0000000000C}

In this Appendix, we prove Eq.~\eqref{eq:appendixA0002}, which in turn holds for any time-dependent, nonsingular, full-rank density matrix ${\rho_t}\in\mathcal{S}$. In this setting, the time derivative of the von Neumann entropy $S({\rho_t}) = -\text{Tr}({\rho_t}\ln{\rho_t})$ is given by
\begin{equation}
\label{eq:appendixC0001}
\frac{d}{dt}S({\rho_t}) = -\text{Tr}\left(\ln{\rho_t}\frac{d{\rho_t}}{dt}\right) - \text{Tr}\left({\rho_t}\frac{d}{dt}\ln{\rho_t}\right) ~.
\end{equation}
Next, we consider the integral representation of the lo\-ga\-rithm of the density matrix $\rho_t$ as follows [see Eq.~\eqref{eq:appendixB0001}]
\begin{equation}
\label{eq:appendixC0002}
\ln{\rho_t} = {\int_0^{\infty}} du \left(\frac{1}{1 + u}\mathbb{I} - {({\rho_t} + u\mathbb{I})^{-1}}\right) ~.
\end{equation}
By taking the time derivative of the previous equation, one gets
\begin{equation}
\label{eq:appendixC0003}
\frac{d}{dt}\ln{\rho_t} = {\int_0^{\infty}} du \, {({\rho_t} + u\mathbb{I})^{-1}}\frac{d{\rho_t}}{dt}{({\rho_t} + u\mathbb{I})^{-1}} ~,
\end{equation}
where we have used that $d{\varrho_t^{-1}}/dt = - {\varrho_t^{-1}}(d{\rho_t}/dt){\varrho_t^{-1}}$, with ${\varrho_t} = {\rho_t} + u\mathbb{I}$ being a nonsingular matrix, i.e., ${\varrho_t^{-1}}{\varrho_t} = {\varrho_t}{\varrho_t^{-1}} = \mathbb{I}$. Hence, by combining Eqs.~\eqref{eq:appendixC0002} and~\eqref{eq:appendixC0003}, one gets
\begin{align}
\label{eq:appendixC0004}
&\text{Tr}\left({\rho_t}\frac{d}{dt}\ln{\rho_t} \right) =\nonumber\\
&= {\int_0^{\infty}} du \, \text{Tr}\left({\rho_t}{({\rho_t} + u\mathbb{I})^{-1}}\frac{d{\rho_t}}{dt}{({\rho_t} + u\mathbb{I})^{-1}}\right) \nonumber\\
&= \text{Tr}\left({\rho_t}\frac{d{\rho_t}}{dt} {\int_0^{\infty}} du \, {({\rho_t} + u\mathbb{I})^{-2}}\right) ~,
\end{align}
where we have applied the identity ${\rho_t}{({\rho_t} + u\mathbb{I})^{-1}} = {({\rho_t} + u\mathbb{I})^{-1}}{\rho_t}$, and also used the cyclic property of the trace. We consider the spectral decomposition $\rho_t = {\sum_l}{p_l}|{l}\rangle\langle{l}|$, with $\langle{j}|{l}\rangle = {\delta_{j,l}}$, while $0 \leq {p_l} \leq 1$ and ${\sum_l}\,{p_l} = 1$. In this case, one gets
\begin{align}
\label{eq:appendixC0005}
{\int_0^{\infty}} du \, {({\rho_t} + u\mathbb{I})^{-2}} &= {\sum_l} {\int_0^{\infty}} \frac{du}{({p_l} + u)^2} \, |{l}\rangle\langle{l}| \nonumber\\
&= {\sum_l}\,{p_l^{-1}}|{l}\rangle\langle{l}| \nonumber\\
&= {\rho_t^{-1}} ~.
\end{align}
By substituting Eq.~\eqref{eq:appendixC0005} into Eq.~\eqref{eq:appendixC0004}, one gets
\begin{align}
\label{eq:appendixC0006}
\text{Tr}\left({\rho_t}\frac{d}{dt}\ln{\rho_t}\right) &= \text{Tr}\left({\rho_t}\frac{d{\rho_t}}{dt}{\rho_t^{-1}}\right) \nonumber\\
&= \text{Tr}\left(\frac{d{\rho_t}}{dt}\right) \nonumber\\
&= 0 ~,
\end{align}
which follows from the fact that $\text{Tr}({\rho_t}) = 1$, where we have also used the cyclic property of the trace. Therefore, Eq.~\eqref{eq:appendixC0006} implies that the time derivative of the von Neumann entropy yields
\begin{equation}
\label{eq:appendixC0007}
\frac{d}{dt}S({\rho_t}) = - \text{Tr}\left(\ln{\rho_t}\frac{d{\rho_t}}{dt}\right) ~,
\end{equation}
and we conclude the proof accordingly.


\section{Proof of Eq.~$\eqref{eq:appendixA0003}$}
\label{sec:0000000000D}

In this Appendix, we prove Eq.~\eqref{eq:appendixA0003}, thus applying for any time-dependent, nonsingular, full-rank density matrix ${\rho_t}\in\mathcal{S}$. We consider the integral representation in Eq.~\eqref{eq:appendixC0003} of the logarithm of the density matrix. Next, we evaluate the following
\begin{align}
\label{eq:appendixD0001}
&\left|\frac{d}{dt}\text{Tr}({\rho_0}\ln{\rho_t})\right| \leq \nonumber\\
&{\int_0^{\infty}} du \, \left|\text{Tr}\left[{({\rho_t} + u\mathbb{I})^{-1}}{\rho_0}{({\rho_t} + u\mathbb{I})^{-1}}\frac{d{\rho_t}}{dt}\right]\right| ~,
\end{align}
where we have used that $\left|\int du {f(u)}\right| \leq {\int} du |{f(u)}|$, and also applied the cyclic property of the trace. By using the matrix norm inequalities $|\text{Tr}({A_1}{A_2}{A_3}{A_4})| \leq {\| {A_1}{A_2}{A_3}\|_{\infty}}{\| {A_4}\|_1} \leq {\|{A_1}\|_{\infty}}{\|{A_2}\|_{\infty}}{\|{A_3}\|_{\infty}}{\| {A_4}\|_1}$, one gets the result
\begin{align}
\label{eq:appendixD0002}
&\left|\frac{d}{dt}\text{Tr}({\rho_0}\ln{\rho_t})\right| \leq \nonumber\\
&{\|{\rho_0}\|_{\infty}}{\left\| \frac{d{\rho_t}}{dt} \right\|_1} \, {\int_0^{\infty}} du \, {\|{({\rho_t} + u\mathbb{I})^{-1}}\|_{\infty}^2} ~.
\end{align}
Next, by using that ${\|{\rho_0}\|_{\infty}} = {\kappa_{\text{max}}}({\rho_0})$, and also ${\|{({\rho_t} + u\mathbb{I})^{-1}}\|_{\infty}} = {({\kappa_{\text{min}}}({\rho_t}) + u)^{-1}}$, where ${\kappa_{\text{min/max}}}(\bullet)$ sets the smallest/largest eigenvalue of the referred density matrix, respectively, we obtain the result
\begin{align}
\label{eq:appendixD0003}
&\left|\frac{d}{dt}\text{Tr}({\rho_0}\ln{\rho_t})\right| \leq \nonumber\\
& {\kappa_{\text{max}}}({\rho_0}) {\left\| \frac{d{\rho_t}}{dt} \right\|_1} \, {\int_0^{\infty}} \frac{du}{({\kappa_{\text{min}}}({\rho_t}) + u)^{2}} ~.
\end{align}
Finally, by evaluating the integral, one gets
\begin{equation}
\label{eq:appendixD0004}
\left| \frac{d}{dt}\text{Tr}({\rho_0}\ln{\rho_t}) \right| \leq \frac{{\kappa_{\text{max}}}({\rho_0})}{{\kappa_{\text{min}}}({\rho_t})} \, {\left\|\frac{d{\rho_t}}{dt}\right\|_1} ~,
\end{equation}
which in turn depends on the Schatten speed, thus concluding our proof.


\section{Quantum relative entropy for single-qubit states}
\label{sec:0000000000E}

In this Appendix, we derive a general formula for the quantum relative entropy for the single-qubit states ${\rho_1} =(1/2)\left(\mathbb{I} + {\vec{r}_1}\cdot\vec{\sigma}\right)$, and ${\rho_2} =(1/2)\left(\mathbb{I} + {\vec{r}_2}\cdot\vec{\sigma}\right)$, where $\vec{r}_1 = {r_1}{\hat{r}_1}$ and $\vec{r}_2 = {r_2}{\hat{r}_2}$ are three-dimensional Bloch vectors, with $\vec{\sigma} = \left({\sigma_x},{\sigma_y},{\sigma_z}\right)$ being the vector of Pauli matrices, and $\mathbb{I}$ is the $2\times2$ identity matrix. By using the integral representation of the logarithm of the density matrix, one gets [see Eq.~\eqref{eq:appendixB0001}]
\begin{align}
\label{eq:appendixE0001}
S({\rho_1}\|{\rho_2}) &= \text{Tr}[{\rho_1}(\ln{\rho_1} - \ln{\rho_2})] \nonumber\\ 
&= {\sum_{l = 1,2}}\, {(-1)^l}\,{\int_0^{\infty}} du \, \text{Tr}[{\rho_1}{({\rho_l} + u\mathbb{I})^{-1}}] ~.
\end{align}
Next, it can be readily shown that~\cite{PhysRevA.91.042330}
\begin{align}
\label{eq:appendixE0002}
{({\rho_l} + u\mathbb{I})^{-1}} &= 2\, {[(1 + 2u)\mathbb{I} + {\vec{r}_l}\cdot\vec{\sigma}]^{-1}} \nonumber\\
&= \frac{2\, [(1 + 2u)\mathbb{I} - {\vec{r}_l}\cdot\vec{\sigma}]}{{(1 + 2u)^2} - {r_l^2}} ~,
\end{align}
for all $l = \{1,2\}$. By combining Eqs.~\eqref{eq:appendixE0001} and~\eqref{eq:appendixE0002}, we arrive at the result
\begin{align}
\label{eq:appendixE0003}
S({\rho_1}\|{\rho_2}) &= 2\,{\sum_{l = 1,2}}\,{(-1)^l}\left\{ {\int_0^{\infty}} du \, \frac{1 + 2u}{{(1 + 2u)^2} - {r_l^2}} \right.\nonumber\\
& \left. - \text{Tr}[({\vec{r}_l}\cdot\vec{\sigma}){\rho_1}] {\int_0^{\infty}} \frac{du}{{(1 + 2u)^2} - {r_l^2}} \right\} ~,
\end{align}
where we have used that $\text{Tr}({\rho_1}) = 1$. Next, by applying the algebraic identities $\text{Tr}({\vec{r}_l}\cdot\vec{\sigma}) = 0$, $\text{Tr}(\mathbb{I}) = 2$, and also $\text{Tr}[({\vec{r}_1}\cdot\sigma)({\vec{r}_l}\cdot\vec{\sigma})] = 2({\vec{r}_1}\cdot{\vec{r}_l})$, we have that Eq.~\eqref{eq:appendixE0003} becomes
\begin{align}
\label{eq:appendixE0004}
S({\rho_1}\|{\rho_2}) &= 2\,{\sum_{l = 1,2}}\,{(-1)^l} \left\{{\int_0^{\infty}} du \, \frac{1 + 2u}{{(1 + 2u)^2} - {r_l^2}} \right. \nonumber\\
&\left. - ({\vec{r}_1}\cdot{\vec{r}_l}) {\int_0^{\infty}}\frac{du}{{(1 + 2u)^2} - {r_l^2}} \right\} ~.
\end{align}
Next, by computing the integrals and proceeding with some algebraic manipulations, it follows that the quantum relative entropy for arbitrary single-qubit states is given by
\begin{align}
\label{eq:appendixE0005}
S({\rho_1}\|{\rho_2}) &= \frac{1}{2}\,{\sum_{l = 1,2}}\,{(-1)^{l + 1}} \left\{\ln\left(1 - {r_l^2}\right) \right. \nonumber\\
&\left. + {r_1}({\hat{r}_1}\cdot{\hat{r}_l})\ln\left(\frac{1 + r_l}{1 - r_l}\right) \right\} ~,
\end{align}
where we have used the following expressions
\begin{equation}
\label{eq:appendixE0006}
{\int_0^{\infty}}\,\frac{du}{{(1 + 2u)^2} - {r_l^2}} = \frac{1}{4{r_l}}\ln\left(\frac{1 + r_l}{1 - r_l}\right) ~,
\end{equation}
and also
\begin{equation}
\label{eq:appendixE0007}
{\int_0^{\infty}} du \,\frac{1 + 2u}{{(1 + 2u)^2} - {r_l^2}} = - \frac{1}{4}\ln\left(1 - {r_l^2}\right) ~.
\end{equation}
In detail, note that Eq.~\eqref{eq:appendixE0005} can be also written as
\begin{align}
\label{eq:appendixE0008}
S({\rho_1}\|{\rho_2}) &= \frac{1}{2}\ln\left(\frac{1 - {r_1^2}}{1 - {r_2^2}}\right) + \frac{r_1}{2}\left[\ln\left(\frac{1 + {r_1}}{1 - {r_1}}\right) \right.\nonumber\\
&\left. - \left({\hat{r}_1}\cdot{\hat{r}_2}\right)\ln\left(\frac{1 + {r_2}}{1 - {r_2}}\right)\right] ~.
\end{align}
Based on Eq.~\eqref{eq:appendixE0008}, it follows that the quantum relative entropy $S({\rho_2}\|{\rho_1})$ reads
\begin{align}
\label{eq:appendixE0009}
S({\rho_2}\|{\rho_1}) &= -\frac{1}{2}\ln\left(\frac{1 - {r_1^2}}{1 - {r_2^2}}\right) + \frac{r_2}{2}\left[\ln\left(\frac{1 + {r_2}}{1 - {r_2}}\right) \right.\nonumber\\
&\left. - \left({\hat{r}_1}\cdot{\hat{r}_2}\right)\ln\left(\frac{1 + {r_1}}{1 - {r_1}}\right)\right] ~.
\end{align}
In particular, for two maximally distinguishable single-qubit states with ${\hat{r}_1}\cdot{\hat{r}_2} = 0$, Eqs.~\eqref{eq:appendixE0008} and~\eqref{eq:appendixE0009} become $S({\rho_{1,2}}\|{\rho_{2,1}}) = (1/2)\ln[(1 - {r_{1,2}^2})/(1 - {r_{2,1}^2})] + ({r_{1,2}}/2)\ln[(1 + {r_{1,2}})/(1 - {r_{1,2}})]$.

Importantly, Eqs.~\eqref{eq:appendixE0008} and~\eqref{eq:appendixE0009} clearly show that the quantum relative entropy is asymmetric for arbitrary single-qubit states, i.e., $S({\rho_1}\|{\rho_2}) \neq S({\rho_2}\|{\rho_1})$. However, Eqs.~\eqref{eq:appendixE0008} and~\eqref{eq:appendixE0009} are expected to recover the same result for the particular case where states $\rho_1$ and $\rho_2$ are related to each other by a unitary evolution. To see this point, suppose that ${\rho_2} = V{\rho_1}{V^{\dagger}} = (1/2)(\mathbb{I} + \vec{v}\cdot\vec{\sigma})$, where ${V^{\dagger}} = {V^{-1}}$ is a unitary matrix, with $V{V^{\dagger}} = {V^{\dagger}}V = \mathbb{I}$. Here, $\vec{v} = {r_1}\hat{v}$, where $\hat{v}$ is a unit vector. In words, the unitary transformation is expected to rotate the probe state (i.e., by changing the direction of unit vector $\vec{r}_1$ in the Bloch sphere), but does not change the purity degree of the quantum state. Hence, we have that $S({\rho_1}\|V{\rho_1}{V^{\dagger}}) = S(V{\rho_1}{V^{\dagger}}\|{\rho_1}) = ({r_1}/2)(1 - {\hat{r}_1}\cdot\hat{v})\ln[(1 + {r_1})/(1 - {r_1})]$.



%

\end{document}